\documentclass{aastex63}

\usepackage{graphicx}
\usepackage{amsmath}

\usepackage{epsfig}

\usepackage{subfigure}
\usepackage{longtable}
\usepackage{times}

\UseRawInputEncoding

\shorttitle{The Rotating Black Hole in$ 4D$ EGB Gravity}
\shortauthors{O.Donmez}

\begin{document}

\title{Dynamical Evolution of the Shock Cone around $4D$ Einstein-Gauss
        Bonnet Rotating Black Hole}
      
\author{O. Donmez}
\altaffiliation{College of Engineering and Technology, American
  University of the Middle East, Kuwait}

\begin{abstract}

  In this paper, a Bondi-Hoyle accretion onto the rotating black hole in Einstein-Gauss Bonnet
  gravity is studied. By injecting the gas from upstream region of the computational
  domain, we have found occurrence of the stable shock cones in the downstream region.
  The dynamical structures and oscillation properties of these shock cones
  strongly depend on the black hole spin parameter $a$ and Gauss-Bonnet coupling
  constant $\alpha$. It is found that the various values of $\alpha$ can lead
  the different amounts of  matter to pile up close to the black hole horizon,  
  higher $\alpha$ causes  bigger oscillation amplitude in the mass
  accretion rate, and the required time to  reach the steady state is getting
  smaller with the increasing in $\alpha$.
  Moreover, increasing $\alpha$ in the negative direction causes
  a decrease in the shock opening angle and  this angle  slightly increases with the increasing
 $\alpha$ in the positive direction. We found that the negative values of
  Gauss-Bonnet coupling constant are more favored to have interesting physical
  outcomes such as accretion rate and oscillation.  
  In addition, the higher the black hole  rotation parameter
  $a$ emerges  the higher the accretion rate. It is also confirmed that, for
  $\alpha \rightarrow 0$,  the black hole solution in EGB gravity converges
  to Kerr in general relativity.
  Furthermore, Gauss-Bonnet  coupling constant could be used to constrain
  the size of  observed shadow of $M87^*$ radius for various values of
  black hole rotation parameter.
  
\end{abstract}



\section{Introduction}
\label{Introduction}

Understanding the nature of strong gravity using the observational and theoretical
tools is still an ongoing process. After  the general theory of the
relativity was formulated by Einstein, many alternative theories  of the
gravity were proposed. The observation of $M87^*$ super-massive
black hole by the Event Horizon Telescope $(EHT)$ collaboration was the first direct evidence
of the strong gravitational regime\citep{Akiyama1, Akiyama2, Akiyama3, Akiyama4, Akiyama5}.
The observed $M87^*$ black hole shadow  indicates that the observed data is well
consistent with the prediction of the general theory of relativity and
the black hole parameters were estimated from the shadow\citep{Kumar2}. On the other hand
the observed shadow has opened a window to have a deep understanding  of the strong
gravity not only in the general relativity but also in alternative theories of gravity.
The alternative theories prove the existing of Kerr black hole and  might be used to
constrain the parameter space of the black hole using EHT collaboration data
\citep{Kumar1, Shaikh1, Bambi1}.

A wind accretion from the $X-$ray binary is an important physical phenomena to reveal
properties of the black hole such as its spin and mass. The simplest accretion scenario
occurring in the $X-$ray binaries  is called Bondi-Hoyle-Lyttleton (BHL) accretion
\citep{Bondi1}. BHL is one of the accretions  studied during many decades using the
tools in Newtonian and general relativistic hydrodynamics. Using the Newtonian hydrodynamics,
firstly, the numerical simulation in $2D$ for an axisymmetric accretion flow
was  performed  for adiabatic gas by \cite{Hunt1}. Later, the analytic study of the accretion flow
onto the compact object and estimation of the accretion rate were computed by \cite{Davies1}.
There were other studies in $2D$ and $3D$ accomplished  recently
\citep{Foglizzo1, Blondin1, MacLeod1, Ohsugi1, Wenrui1}. BHL accretion around
the black hole was extensively studied using the general relativistic hydrodynamics and
magneto-hydrodynamics either in case of spherical symmetry or axial symmetry
\citep{Donmez6, Penner1, Donmez5, Penner2, LoraClavijo1,  LoraClavijo2, CruzOsorio1}.
Studying BHL accretion onto the rotating black hole using the modified gravity may
specify more information about the rotating black hole.

A newly discovered $4D$ Einstein-Gauss-Bonnet (EGB) gravity opened a new  window
to define a black hole in EGB gravity.  This new discovery, which contains
static and spherically symmetric black hole,
was used to extract important features of the astrophysical phenomena in different aspect.
Number of work have been done to reveal the properties of the astrophysical system.
Physical properties of the black hole \citep{Roman1, Rittick1}, 
the gravitational lensing by the black hole \citep{Islam1}, Hawking radiation of the
massless scalar \citep{Zhang1}, the observational limits on the Gauss-Bonnet
coupling constant \citep{Feng1, Timothy1}, the radiating black holes
\citep{Sushant1, Ghosh2, Yunlong1}, the last stable circular orbit for photons and
particles \citep{Minyong1, Zhang1}, the non-relativistic matter perturbations  
growth rate \citep{Zahra1} were extensively studied.
In addition, the rotating black hole in EGB gravity were studied to
investigate Gauss-Bonnet coupling constant $\alpha$ on the size of shadow in the
context of massive black hole observation $M87^*$  \citep{Kumar1, Wei1},
to find energy extraction efficiency for particle \citep{Yunlong1}, and
to extract the center of mass energy of the two colliding particles \citep{Kumara1}.

After rescaling of Gauss-Bonnet coupling constant $\alpha$, this constant makes a
non-trivial contribution to the gravitational dynamics in a strong gravitational
region. This important feature of EGB gravity starts to pay attention to understand
the effects of $\alpha$ to different astrophysical problems around the black
hole. In \citet{Donmez3} they explored the properties of the shock cone in a strong
gravitation region around the non-rotating black hole in EGB gravity. The cone was
produced as a consequence of the Bondi-Hoyle accretion and was connected to the black
hole horizon on the downstream side of the computation domain. It was found that the
Gauss-Bonnet coupling constant $\alpha$ plays an important role not only in the creation of
a shock cone but also in its oscillation properties. Increasing $\alpha$ caused
 strong oscillations inside the shock cone. These strong oscillations would lead
to the  Quasi-Periodic Oscillations (QPOs). Besides, they also studied the effects of the
bigger $\alpha$ values in the  negative direction and found that the oscillation
amplitude of the shock cone would be suppressed.

In this paper, we study  the properties of the shock cones and
their oscillation properties in case  of the Bondi-Hoyle accretion
on the equatorial plane around
the rotating black hole in EGB gravity. We numerically
model the Bondi-Hoyle accretion to explore the effects of the black hole rotation
parameter $a$ and Gauss-Bonnet coupling constant $\alpha$ onto the shock cone dynamics.
For this, we have done systematic work to find out the shock cone structures
using different values of $a$ and $\alpha$. In particular, we calculate the mass
accretion rate and QPOs to understand the dynamical structure of the shock cone and
its oscillatory behavior as a function of $a$ and $\alpha$. We have also constrained
the Gauss-Bonnet coupling constant $\alpha$ for the recent discovery, super-massive
black hole $M87^*$. 
in EGB gravity and general relativity. The possible effects of both
constant $(a, \alpha)$ on the oscillation properties of the shock cones are extensively
explored.

The plan of the paper is as follows: The brief summary of recently proposed the
rotating black hole in $4D$ EGB gravity and real solutions for various values of 
$\alpha$ and black hole spin $a$ are summarized in Section
\ref{Non-rotating Black Hole Solution of 4D EGB Gravity}. In Section
\ref{GRHE}, the general form of General Relativistic Hydrodynamical (GRHD)
equations on equatorial plane, initial conditions for primitive variables, 
black hole spin parameter, Gauss-Bonnet coupling constant, and useful formulas to
demonstrate the numerical results are described. In Section \ref{BondiHoyle},
the Bondi-Hoyle accretion onto the rotating black hole in EGB gravity are
extensively presented. The different aspects of the numerical results for
varying values of $\alpha$ and $a$ are discussed computing the mass
accretion rates and the power density spectra inside the shock cones.
In addition, a possible application  of our numerical result to the  
observed  $M87^*$ black hole shadow is speculated. 
The summary and some concluding remarks are given in Section \ref{Conclusion}.
The geometrized unit is studied throughout the paper, $G = c= 1$.


\section{Rotating Black Hole Solution of 4D EGB Gravity}
\label{Non-rotating Black Hole Solution of 4D EGB Gravity}

The non-rotating static black hole solution in $4D$ EGB gravity was defined
by rescaling Gauss-Bonnet coupling constant $\alpha \rightarrow \alpha/(D-4)$ 
in the limit $D \rightarrow 4$ \citet{Glavan1} and it is given as;

\begin{eqnarray}
 ds^2 = -f(r)dt^2 + \frac{1}{f(r)}dr^2 + r^2d\theta^2 + r^2sin(\theta)d\phi^2,
\label{EGB1}
\end{eqnarray}

\noindent
where

\begin{eqnarray}
  f(r) = 1 + \frac{r^2}{2\alpha}\left(1 - \sqrt{1 + \frac{8 \alpha M}{r^3}} \right).
\label{EGB2}
\end{eqnarray}

In order to generate the rotating black hole solution in $4D$ EGB gravity,
the advance null Eddington-Finkelstein coordinates $(u,r,\theta,\phi)$ are used
with an approach \citep{Azreg1} in Eq.\ref{EGB1} \citep{Ghosh1, Wei1}. The used
transformation to define the metric in Eddington-Finkelstein coordinates is

\begin{eqnarray}
  du = dt - \frac{dr}{f(r)}.
\label{EGB3}
\end{eqnarray}

\begin{figure*}
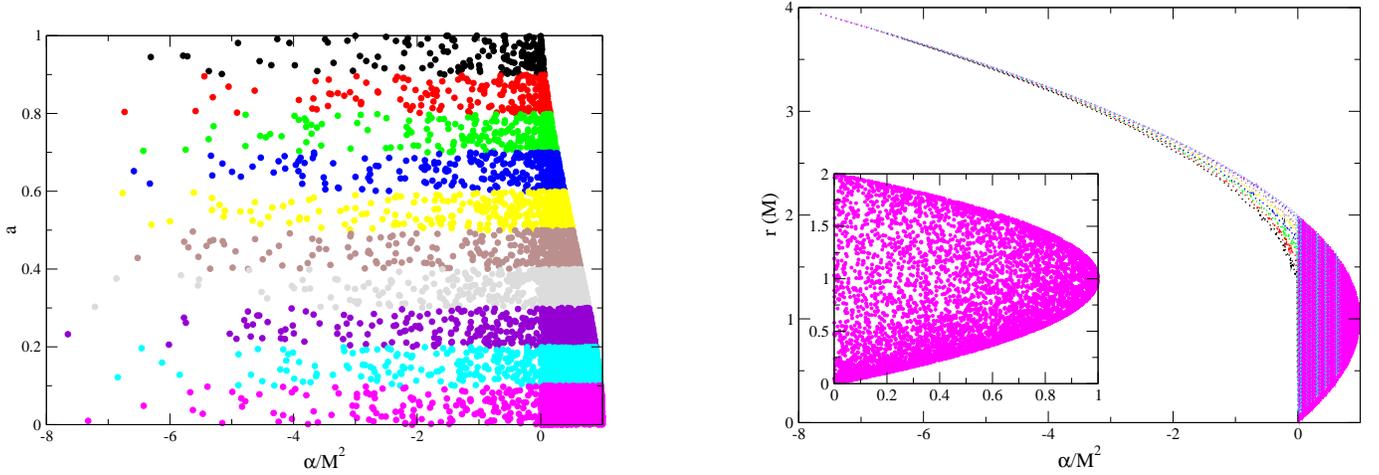

  \vspace{1cm} 
  \center
  \psfig{file=Fig01.eps,width=8.0cm}
  \hfill
    \psfig{file=Fig02.eps,width=8.0cm}
    \caption{The existence of the black hole in EGB gravity.
      {\bf Left panel:} Each point represents the parameter space $(a, \alpha)$ of the black hole.
      A less number of real solutions occur for the higher values of $\alpha$ in the negative
      direction for varying black hole spins. {\bf Right panel:} It shows the regions
      for an existence of black hole space-time for various $\alpha$ with either
      degenerate horizon radius or two distinct ones.
 }
\label{EGB_R1}
\end{figure*}

After the statically symmetric black hole metric can be written in the
advanced null coordinates, the set of the null tetrad is introduced \citep{Ghosh1, Wei1}.
Then radial coordinate $r$ is defined in the complex form in 
the modified Newman-Janis algorithm. On the other hand, the metric functions can be represented
with undefined ones which are $f(r) \rightarrow F(r,a,\theta)$ and $r^2 \rightarrow H(r,a,\theta)$.
Using the transformation and finding new null tetrads, the rotating black hole metric in the 
Eddington-Finkelstein coordinates is given by \citep{Ghosh1, Wei1}

\begin{eqnarray}
  ds^2 &=& -Fdu^2 - 2dudr + 2asin^2\theta(F-1)dud\phi + 2asin^2\theta drd\phi +
  H d\theta^2 + 
  sin^2\theta(H + a^2sin^2\theta (2 -F))d\phi^2,
  \label{EGB4}
\end{eqnarray}

\noindent
where $a$ is dimensionless black hole spin parameter. Using the global coordinates
$du = dt' + \lambda(r)dr$ and $d\phi = d\phi' + \chi(r)dr$,  Eq.\ref{EGB4}
can be written in Boyer-Lindquist coordinates. The function seen in global coordinates
are $\lambda(r) = -\frac{r^2+a^2}{f(r)r^2+a^2}$ and
$\chi(r) = -\frac{a}{f(r)r^2+a^2}$ \citep{Azreg1}. The undefined functions are obtained as
$F = \frac{f(r)r^2+a^2cos^2\theta}{H}$ and $H = r^2 + a^2cos^2 \theta$. Finally,
the metric for the rotating black hole in EGB gravity is

\begin{eqnarray}
  ds^2 &=& -\frac{\Delta - a^2sin^2\theta}{\Sigma}dt^2 + \frac{\Sigma}{\Delta}dr^2 -
  2asin^2\theta\left(1- \frac{\Delta - a^2sin^2\theta}{\Sigma}\right)dtd\phi + 
  \Sigma d\theta^2 + \nonumber \\
  && sin^2\theta\left[\Sigma +  a^2sin^2\theta \left(2- \frac{\Delta -  
   a^2sin^2\theta}{\Sigma} \right)  \right]d\phi^2,
\label{EGB5}
\end{eqnarray}

\noindent
where $\Sigma$ and $\Delta$ read as,

\begin{eqnarray}  
  \Sigma &=& r^2 + a^2cos^2\theta \nonumber \\
  \Delta &=& r^2 + a^2 + \frac{r^4}{2\alpha}\left(1 - \sqrt{1 + \frac{8 \alpha M}{r^3}} \right),
 \label{EGB6}
\end{eqnarray}

\noindent
where $a$, $\alpha$, and $M$ are spin parameter, Gauss-Bonnet coupling constant, and
mass of the black hole, respectively. The horizons of the black holes were obtained
numerically by solving $\Delta=0$  and given in Fig.\ref{EGB_R1}. Each dot on both
figures represents the real solutions for various values of $\alpha$ and black hole spin
$a$.

%



\section{General Relativistic Hydrodynamical Equations, Initial and Boundary Conditions}
\label{GRHE}

The General Relativistic Hydrodynamics (GRH) can be written in terms of conservations
of the mass and energy-momentum

\begin{eqnarray}
  \bigtriangledown_{a}T^{ab}=0,  \nonumber \\
  \bigtriangledown_{a}(\rho u^{a})=0,
\label{GRH1}
\end{eqnarray}

\noindent where $T^{ab}$ is the stress-energy-momentum tensor, $u^{a}$ is the four-velocity of the
fluid, and $\rho$ is the rest-mass density of the fluid. The stress-energy-momentum tensor
for prefect fluid is

\begin{eqnarray}
 T^{ab} = \rho h u^{a}u^{b} + P g^{ab},
\label{GRH2}
\end{eqnarray}

\noindent where enthalpy is $h = 1 + \epsilon + \frac{P}{\rho}$. $\epsilon$, $P$, and $g^{ab}$
are the specific internal energy, pressure, and the inverse of the space-time metric,
respectively.

The flux-conserving form of Eq.\ref{GRH1} can be written using $3+1$ formalism which are

\begin{eqnarray}
  \partial_t\left(\sqrt{\gamma}W\rho\right) +   \partial_i \left((\tilde{\alpha}v^i -
  \beta^i)\sqrt{\gamma}W\rho\right)  &=& 0\nonumber \\
  \partial_t\left(\sqrt{\gamma}\rho W^2 h v_j\right) + \partial_i \left((\tilde{\alpha}v^i -
  \beta^i)\sqrt{\gamma}\rho W^2 h v_j +
  \tilde{\alpha} \sqrt{\gamma}P\delta_{j}^{i}\right)  &=& \tilde{\alpha} \sqrt{\gamma}T^{ab}g_{bc}\Gamma^{c}_{aj} \nonumber \\
  \partial_t\left(\tau \right) +
  \partial_i \left((\tilde{\alpha}v^i - \beta^i)\tau + \tilde{\alpha}\sqrt{\gamma}P v^i\right) &=&
  \tilde{\alpha}\sqrt{\gamma}(T^{a0}\partial_a \tilde{\alpha} - \tilde{\alpha}T^{ab}\Gamma^{0}_{ab}),
\label{GRH3}
\end{eqnarray}

\noindent where the energy conserved variable $\tau =\sqrt{\gamma}\left(\rho h W^2 - P -W \rho\right)$,
$\partial_t =  \frac{\partial}{\partial t}$, and $\partial_i = \frac{\partial}{\partial x^i}$.
$v^i$ is three-velocity of the fluid.  The indices $a$, $b$, $c$,  and $d$ run from $0$ to $3$ and
Latin indices $i$ and $j$ run from $1$ to $3$. The $4-$ dimensional Christoffel symbol is
$\Gamma^{c}_{ab} = \frac{1}{2}g^{cd}\left(\partial_a g_{bd} +  \partial_b g_{ad} - \partial_d g_{ab}\right)$.
The $4-$ velocity components are related to $3-$velocity with the following expression
$u^i = W(v^i - \beta^i/\tilde{\alpha})$. The four metric $g_{ab}$, its inverse $g^{ab}$,
Christoffel symbol $\Gamma^{c}_{ab}$, lapse function $\tilde{\alpha}$, and shift vector $\beta^i$
are defined on the equatorial plane by using the metric for the rotating black
hole in EGB gravity given in Eq.\ref{EGB5}.

The lapse function $\tilde{\alpha}$ is,

\begin{eqnarray}
  \tilde{\alpha} = \sqrt{\frac{a^2(1-f(r))^2}{r^2+a^2(2-f(r))} + f(r)}
\label{GRH4}
\end{eqnarray}

\noindent
where $f(r)$ is given in Eq.\ref{EGB2}. And the shift vectors are,

\begin{eqnarray}
  \beta_r &=& 0, \nonumber \\
  \beta_{\phi} &=& \frac{a r^2}{2\pi \alpha}\left(1 -\sqrt{1 + \frac{8 \pi \alpha M}{r^3}}\right),  \nonumber \\
  \beta_{\theta} &=& 0. 
\label{GRH5}
\end{eqnarray}

The High Resolution
Shock Capturing (HRSC) scheme is used to solve Eq.\ref{GRH3} along with
Marquina fluxes, and MUSCL left and right states of the primitive
variables at each cell center \citep{Donmez1, Donmez2, Donmez5}.

\begin{table}
\footnotesize
\caption{$\alpha$ is Gauss-Bonnet coupling constant,
  $a$ is the dimensionless  black hole rotation parameter,
  $r_{in}$ is the inner radius of computational domain,
  $t_{s}$ (saturation time) is the time to reach the quasi-steady state,
  and $\Delta \Phi = \lvert A_{\max}-A_{min} \rvert$ measures the distance
between a crest and a through.}
 \label{Inital_Con}
\begin{center}
  \begin{tabular}{ccccc}
    \hline
    \hline
 $\alpha (M^2)$ & $a$ & $r_{in} (M)$ & $t_{s} (M)$ & $\Delta \Phi$ \\
 \hline
 $0.253$   & $0.7$   & $2.0$ & $\sim 1088$ & $24.48$ \\
 $0.31599$ & $0.644$ & $1.7$ & $\sim 1235$ & $17.91$ \\ 
 $0.757$   & $0.28$ & $2.0$ & $\sim 1086$ & $43.37$ \\
 \hline
 $-5.16$   & $0.9$   & $3.7$ & $\sim 937$ & $30.2$  \\
 $-1.691$  & $0.768$ & $2.7$ & $\sim 1288$ & $27.8$  \\
 $-0.345$   & $0.9$   & $2.1$ & $\sim 1009$ & $24.2$  \\
 $0.000625$& $0.9$   & $1.9$ & $\sim 996$ & $27.8$  \\
 $0.054$   & $0.9$   & $1.8$ & $\sim 995$ & $24.8$  \\
 \hline
 $0.9997$  & $0.0048$& $1.7$ & $\sim 1510$ & $39.96$  \\
 \hline
 $-0.422$  & $0.952$ & $2.2$ & $\sim 900$ & $20.89$  \\
 $-2.924$  & $0.967$ & $3.2$ & $\sim 958$ & $34.5$  \\ 
 $-4.9041$ & $0.616$ & $3.7$ & $\sim 978$ & $34.5$  \\
 \hline
 $-4.93$   & $0.28$  & $3.7$ & $\sim 1415$ & $29.72$  \\
 $-3.03$   & $0.28$  & $3.7$ & $\sim 1338$ & $40.28$  \\
 $-0.99$   & $0.28$  & $3.7$ & $\sim 1307$ & $41.88$  \\
 $-0.37$   & $0.28$  & $3.7$ & $\sim 1159$ & $38.19$  \\
 $0.096$   & $0.28$  & $3.7$ & $\sim 1093$ & $40.65$  \\
 $0.41 $   & $0.28$  & $3.7$ & $\sim 1073$ & $40.35$  \\
 $0.68 $   & $0.28$  & $3.7$ & $\sim 1039$ & $38.90$  \\
 \hline
 &   &Kerr Black Hole &  &  \\
 \hline
 $-$ & $0.28$ & $2.2$ & $\sim 1140$ & $39.03$   \\
 $-$ & $0.73$ & $2.0$ & $\sim 973$ & $24.97$   \\ 
\hline
\hline
  \end{tabular}
\end{center}
\end{table}

The numerical simulation is performed in polar coordinate $(r,\phi)$
in the vicinity of the black hole at the center, assuming spherical symmetry. The
detailed information about the code can be found in \citet{Donmez1, Donmez2}. To study the
Bondi-Hoyle accretion towards the rotating black hole in EGB gravity, the velocity
components of initial flow at upper boundary are given as

\begin{eqnarray}
  V^r= \sqrt{\gamma^{rr}}V_{\infty}cos(\phi) \nonumber \\
  V^{\phi}= -\sqrt{\gamma^{\phi \phi}}V_{\infty}sin(\phi)
\label{GRH6}
\end{eqnarray}

\noindent
where $V_{\infty}$ represents the asymptotic velocity at infinity.
We let the gas fall towards the black
the black hole homogeneously by choosing these velocities. In order to understand
the effects on  Gauss-Bonnet coupling constant $\alpha$ and black hole spin parameter $a$ onto the
shock cones, we fixed the values of asymptotic velocity $V_{\infty}=0.3$, the sound speed $c_{s,\infty}=0.1$,
and adiabatic index $\Gamma = 4/3$. More detailed information about handling the
initial conditions and other details can also be found in \citet{Donmez3}.

The computation domain is defined on the equatorial plane in polar coordinate and physical
boundaries are located at $r_{min} \leq r \leq 100M (10 r_{acc})$ and at $0 \leq \phi \leq 2\pi$.
The accretion radius is

\begin{eqnarray}
  r_{acc} = \frac{M}{c_{\infty}^2 + V_{\infty}^2}. 
\label{GRH7_1}
\end{eqnarray}

\noindent In order to reduce the effect of the outer boundary, it should be located at least
$4 r_{acc}$. It is at $10 r_{acc}$ in all our numerical simulations.
As seen in Table \ref{Inital_Con},  $r_{min}$ can vary from model to model which is chosen
very close the event horizon of the black  hole. We divided the physical domain to the uniform
cells using $1024$ points along $r$ and $512$ points at $\phi$ directions.

We use the second-order numerical scheme so that we need to define two ghost zones
at the inner and outer boundaries of the computational domain along the $r$. These
zones are filled by copying the corresponding values from the first interior data
inside the physical domain. The adopted boundary along the $\phi$
direction is the periodic boundary condition.

In order to understand the dynamical behavior of accretion disk in the case of Bondi-Hoyle
accretion, the mass accretion rate onto the black hole is computed, assuming that the spherical
detector is located close to the event horizon. The expression of the mass accretion rate is

\begin{eqnarray}
 \frac{dM}{dt} = -\int_0^{2\pi}\tilde{\alpha}\sqrt{\gamma}\rho u^r d\phi.
\label{GRH7}
\end{eqnarray}

\noindent Although the accreated mass into the black hole would increase its
mass a  very small amount,
It is fair to assume that the black hole mass is constant during evolution.

The angular momentum transfer would reveal some features of the shock cone dynamics.
In order to understand the relationship between the rotating accreated matter and
black hole parameters $(a, \alpha)$, we compute the angular momentum flux at the inner
boundary of the domain very close the black hole horizon along the spherical surface.
The azimuthal component of the angular momentum flux on the equatorial plane is

\begin{eqnarray}
 \frac{dL}{dt} = -\int_0^{2\pi}\tilde{\alpha}\sqrt{\gamma}\rho h u^r u^{\phi} d\phi.
\label{GRH8}
\end{eqnarray}

\noindent One of our main goals is to find out how the oscillating shock cone transport the angular
momentum through the disk; radially outward or inward and its dependencies to the Gauss-Bonnet
coupling constant and black hole rotation parameter..

\section{The Accretion onto 4D EGB Rotating Black Hole}
\label{BondiHoyle}

Here, we basically focus on the dynamical evolutions of the shock cones and their oscillation
properties around the rotating black hole in EGB gravity.
The accretion is generated by Bondi-Hoyle accretion injecting gas from upstream region of
the computational domain toward the black hole. The shock cone appears at the downstream region
with a rigid opening angle. In order to extract the physical properties of these cones,
we compute and plot the mass accretion rates, angular momentum accretion rate, shock cone
opening angles, oscillation amplitudes of the matter inside the cone, and power spectrum density
for varies values of Gauss-Bonnet coupling constant $\alpha$ and black hole spinning parameter $a$.
Some of the important parameters used in numerical simulations and extracted from
the calculations are given in Table \ref{Inital_Con}.

\subsection{Numerical Results and Discussion}
\label{Numerical Results}

\begin{figure*}
  \vspace{1cm}
  \center
  \psfig{file=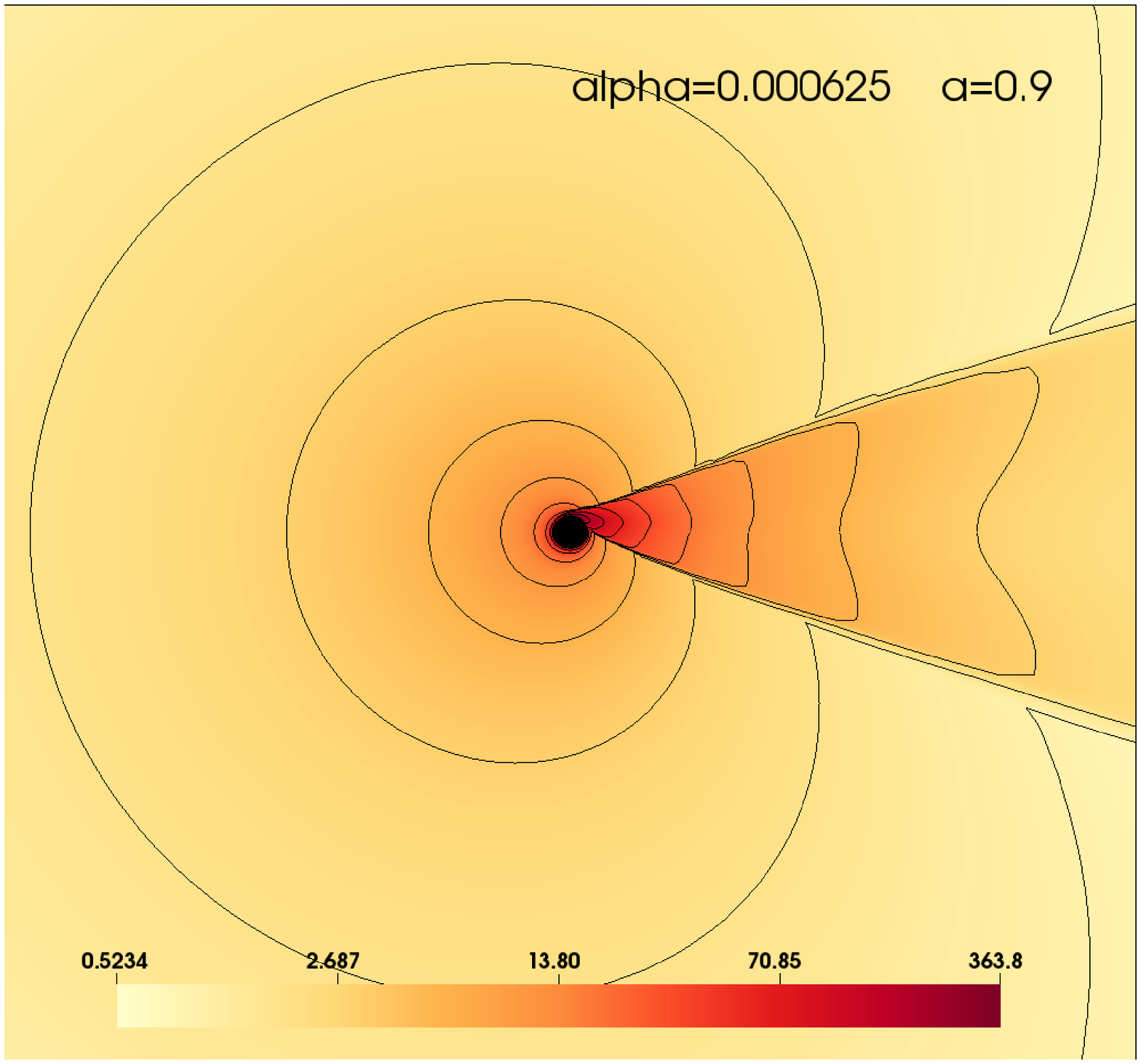,width=8.0cm}
  \psfig{file=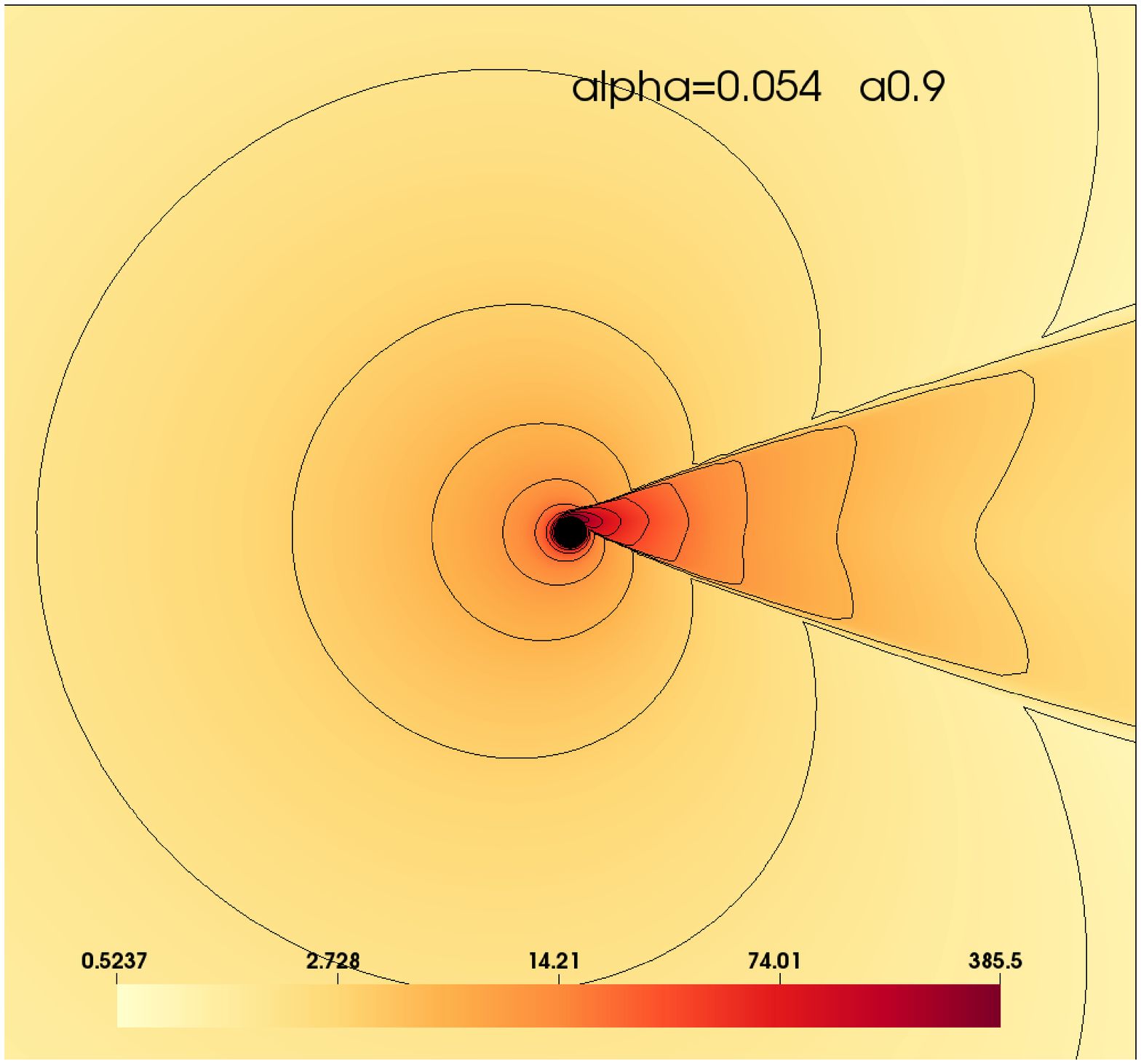,width=8.0cm}
  \psfig{file=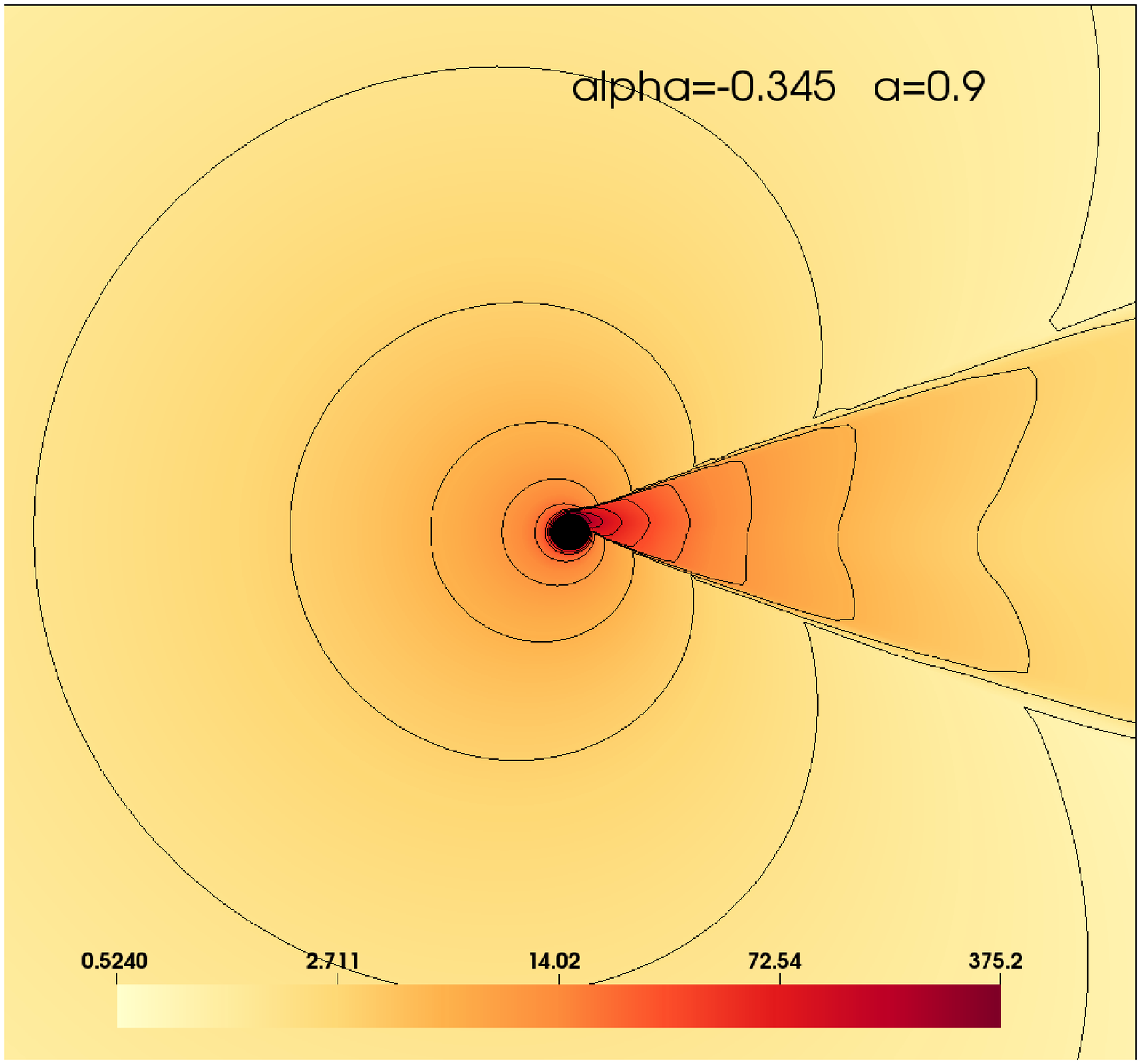,width=8.0cm}
  \psfig{file=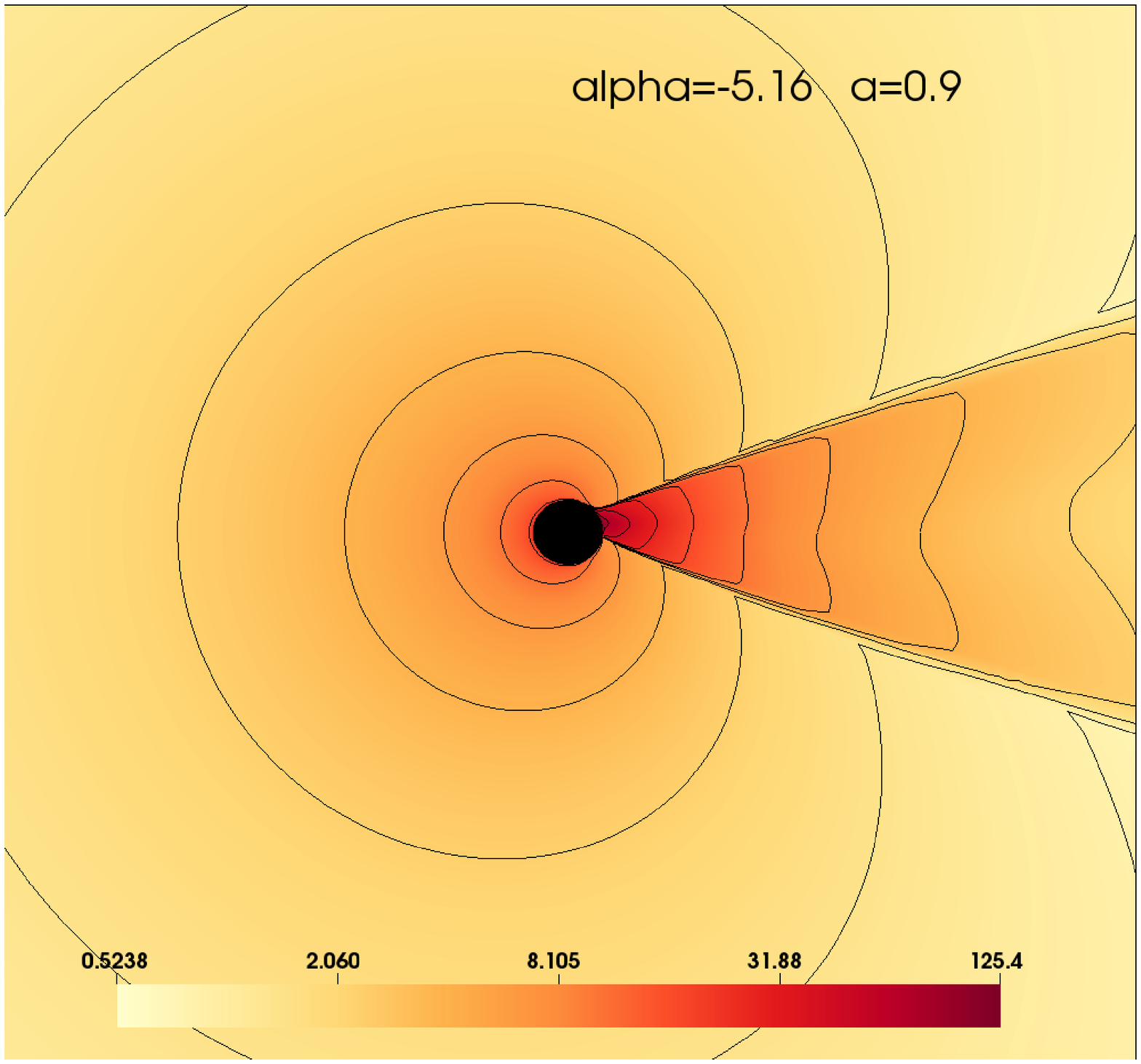,width=8.0cm}
  \caption{The logarithmic rest-mass density on the equatorial plane
    around the rotating black hole $a=0.9$ in EGB gravity. The close-up view of the
    snapshots are plotted at much later than the shock cone reached to the steady-state. The color
    contour plots show the shock cone dynamics for varying values of Gauss-Bonnet coupling
    constant $\alpha$ and the close-up view boundaries are located at $[x,y] \rightarrow [-60M,60M]$.}
\label{NR_00}
\end{figure*}

In Fig.\ref{NR_00}, the color plot of the rest mass-density with their density counter
is shown on the equatorial plane around the rotating black hole $a=0.9$ in EGB gravity
for different values of Gauss-Bonnet coupling constant $\alpha$ at $t \sim 12000 M$,
much later than the time to require to reach the saturation point, $\sim 1000 M$.
The shock cones are formed and bent around the black hole
due to the warped space-time around the rotating black hole. The bended space-time is
more clearly seen in case at which the inner boundary of the computation domain is
more closer to the black hole. The rest-mass density is high at the inner boundary close to the
back hole horizon. As it is seen in left snapshots $\alpha=0.000625$ and $\alpha=-0.345$,
having a higher rest-mass density does not only depend on the
 inner boundary location but it also slightly changes with $\alpha$.
 Even though the inner boundary for  $\alpha=0.000625$ is closer to the horizon
 than $\alpha=-0.345$, the rest mass-density is relatively higher for $\alpha=-0.345$.
 The higher  density  could cause the hotter  shock cone so that we may expect
 the higher energetic phenomena close the black hole horizon for different values of $\alpha$.
 
\begin{figure*}
  \vspace{1cm} 
  \center
  \psfig{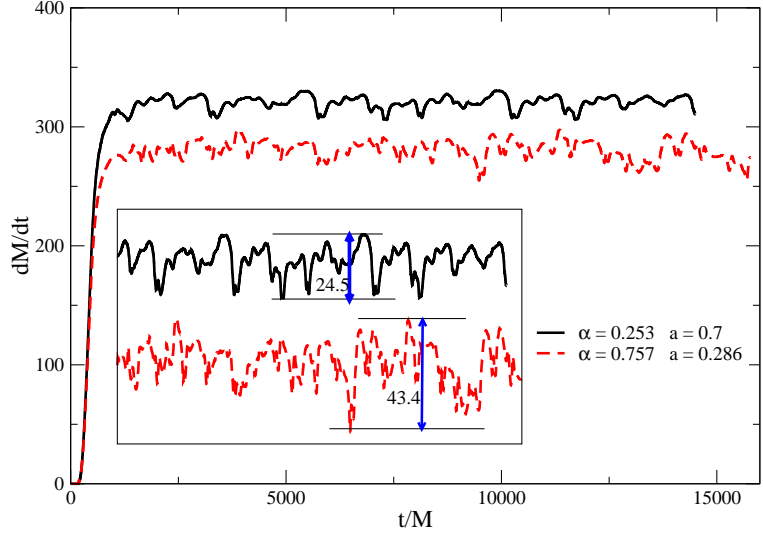}
  \caption{The comparison of the mass accretion rates for two different cases.
    The accretion process reaches a quasi-stationary state steady around $1200 M$.
    While the higher the black hole spin causes the more matter to accreate towards
    the black hole, the oscillation amplitude gets larger for the higher positive
    value of the Gauss-Bonnet coupling constant $\alpha$.}
\label{NR_1}
\end{figure*}

For the different values of Gauss-Bonnet coupling constant $\alpha$
and black hole rotation parameter $a$ with the same initial setup,
in the numerical simulations, the major
differences appear not only in the maximum saturation value of mass accretion rate
but also in the oscillation property. Fig.\ref{NR_1} shows the mass accretion rate
as a function of time for different values of $\alpha$ and $a$. The values of $\alpha$
in first model is almost same as  $a$ in the second model. While larger
$a$ emerges  a higher accretion rate  computed at location $r=6.5M$, 
higher $\alpha$ causes a bigger oscillation amplitude in the mass accretion rate.
The bigger gradient in the mass accretion rate, after the shock cone reaches to the steady state,
would lead to a more chaotic radiation in observed phenomena. 

\begin{figure*}
  \vspace{1cm}   
  \center
  \psfig{file=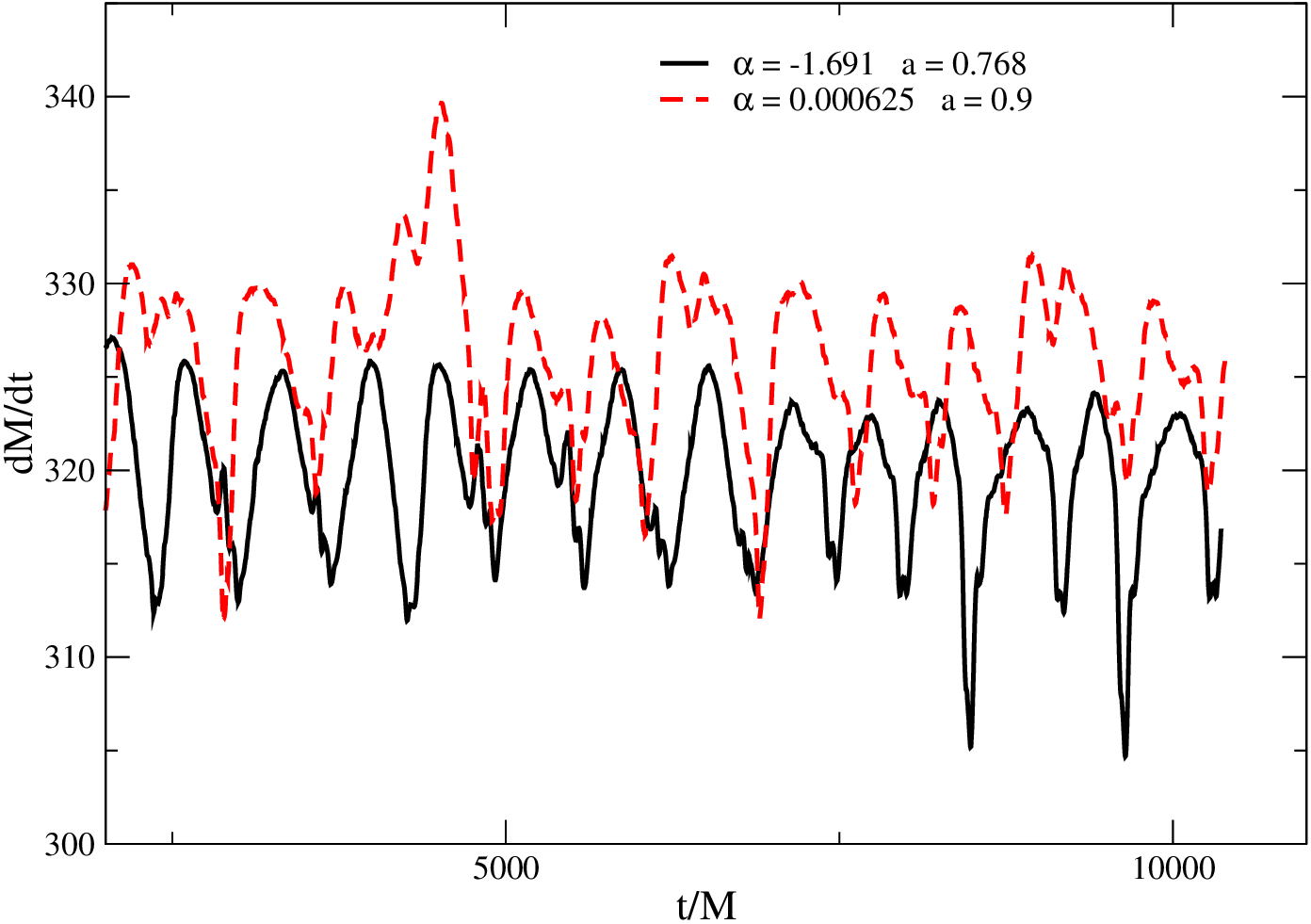,width=10.0cm}
  \caption{The same as Fig.\ref{NR_1} but for different values of $a$ and $\alpha$.
    The distance between a crest and a through is
    $\psi = \lvert A_{\max}-A_{min} \rvert$ = $27.8$ in both model.}
\label{NR_2}
\end{figure*}

In Fig.\ref{NR_2}, the mass accretion rates for  two sets of data are reported.
The rates are plotted after they reach to the steady accretion
state. The mass accretion rate has different dependency on the
Gauss-Bonnet coupling constant $\alpha$ and the black hole rotation parameter $a$.
As it seen from this figure, there is no striking difference between the models
and they show the same type of behavior with a small gradient during the evolution
in given models eventhough the inner radius of these two cases do not equal to each other.
Table \ref{Inital_Con} shows that the black hole shadow radius is
bigger for the model $\alpha=-1.691$  and $a=0.768$.
It might show us that the same physical situation could be reached for
different values of $\alpha$ and $a$. Although the observed shadow for  $M87^*$
\citep{Akiyama1, Akiyama2} is consistent with numerical results in the general relativity,
we still have some open doors to  find a relation between black hole rotation parameter
and  parameters in modified gravity for the observed $M87^*$ \citep{Kumar1}. So that
the black hole shadow diameter could be constrained with Gauss-Bonnet coupling
constant $\alpha$ to get consistency between the numerical simulation and observation.
It is shown from the numerical simulations that the model, $\alpha=-1.691$  and $a=0.768$,
with a smaller black hole rotation parameter and a bigger shadow radius could be used to
understand the physical properties of the observed black hole $M87^*$ by $EHT$.

\begin{figure*}
 \vspace{1cm} 
  \center
  \psfig{file=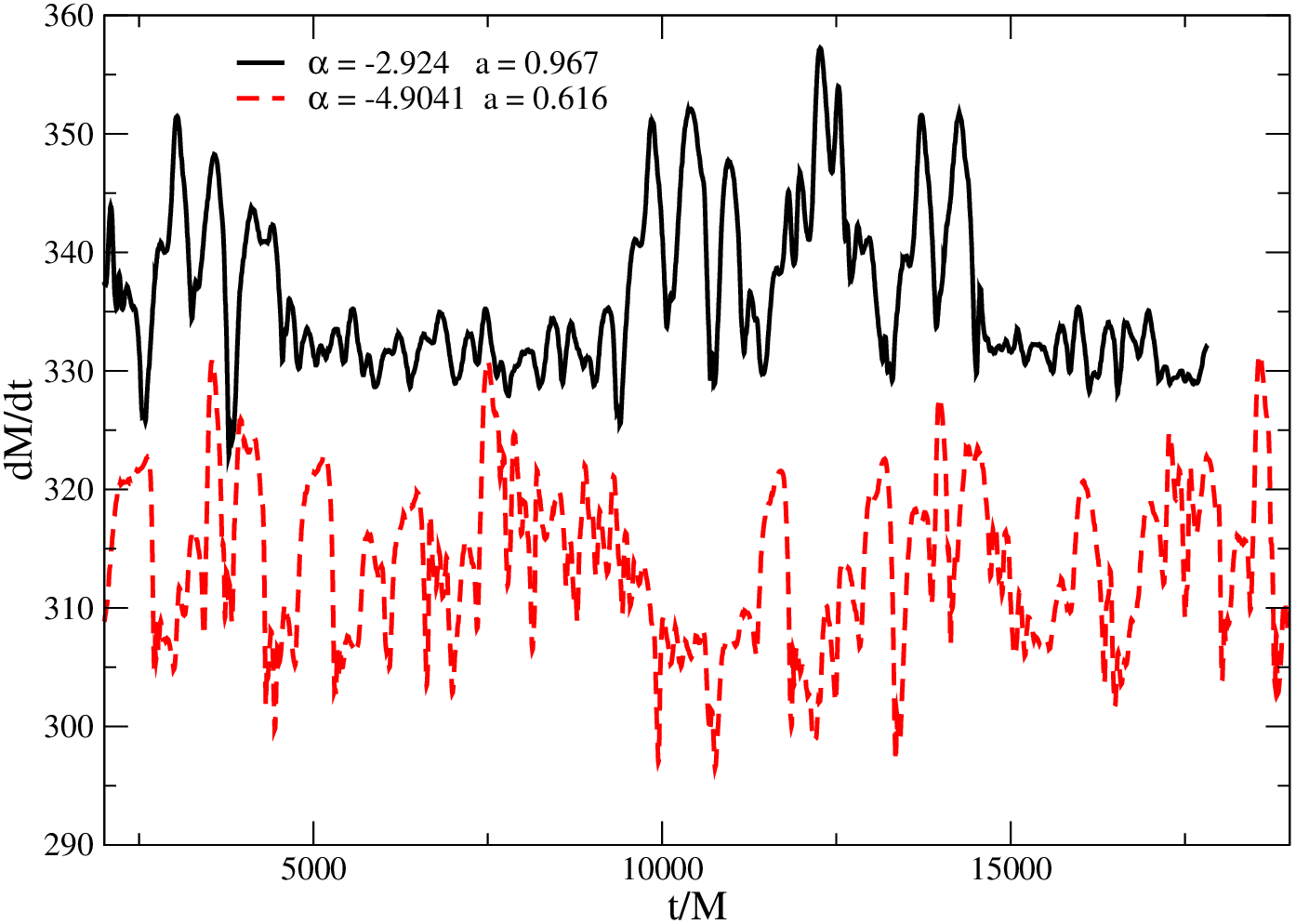,width=10.0cm}
  \caption{The same as Fig.\ref{NR_1} but for different values of $a$ and $\alpha$.
    The distance between a crest and a through is
    $\psi = \lvert A_{\max}-A_{min} \rvert$ = $34.5$ in both model.}
\label{NR_3}
\end{figure*}

In Fig.\ref{NR_3}, the extreme values of Gauss-Bonnet coupling constant $\alpha < -2$  and
the black hole rotation parameter $a > 0.6$ produce more chaotic motions in the mass-accretion
rate when it is compared with the results given in Figs.\ref{NR_1} and \ref{NR_2}. It may be associated
with the effect due to the coupling between fastly rotating black hole parameter
and bigger values of $\alpha$.
Larger the values of these parameters produces larger the oscillation amplitude in mass
accretion rate.  These dependences of mass accretion rate on the black hole parameters
reveal more detail effects and cause rapid growth on it. The results in Figs. \ref{NR_2}
and \ref{NR_3} indicate that the negative values of $\alpha$ is more interesting to study
to extract more details about physical system. It was also put forward by \citet{Wei1} when
they were studying the rotating black hole shadow.

\begin{figure*}
 \vspace{1cm} 
  \center
  \psfig{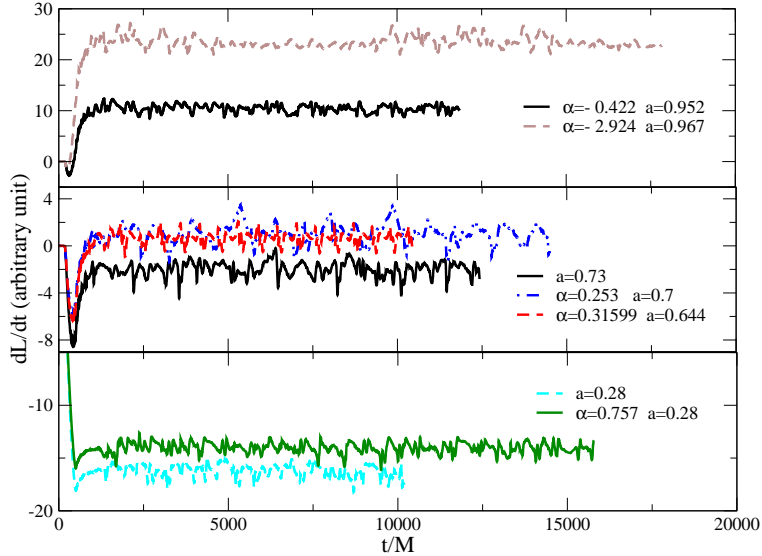}
  \caption{The comparison of the angular momentum fluxes versus time computed
    at the inner boundary of the computational domain for different models.
    The angular momentum transformation strongly depends on the black hole spin and
  Gauss-Bonnet coupling constant $\alpha$.}
\label{NR_4}
\end{figure*}

The Bondi-Hoyle accretion creates a steady-state shock cone around the rotating black hole.
The falling gas gets toward the black hole due to the gravitational force and settles into the
equatorial plane.  The angular momentum of the rotating gas can play an important
role in the moving of gas towards or away from the black hole. In oder to measure the strength
of how the angular momentum would be transferred outwards, we  compute the angular momentum
flux which clearly shows the effect of various values of  Gauss-Bonnet coupling
constant $\alpha$ and the black hole rotation parameter $a$ to the shock cone
dynamics, seen in Fig.\ref{NR_4}. As it is seen in upper part of Fig.\ref{NR_4},
when changing $\alpha$, a difference appears in the strength of angular
momentum flux for the same values of $a$ although oscillation behavior is almost the same.
The higher the $\alpha$ in negative direction leads the bigger in the angular momentum flux,
that is, the more angular momentum would be transferred outward for  $\alpha=-2.924$.
It is also consistent with the energy extraction efficiency which increases with the 
larger value of  $\alpha$ in the negative direction \cite{Yunlong1}.
The angular momentum flux around the Kerr and rotating black holes in EGB
gravity for two different values of $\alpha$ are given in the middle part of Fig.\ref{NR_4}
using the similar rotation parameter $a = 0.7$. The angular momentum flux almost oscillates around
zero value while it is negative for the gas rotating around Kerr black hole.
The result in EGB gravity  definitely shows a deviation from the Kerr solution.  
We have observed the same trend in the lower part of the same figure.
It is clearly seen from the simulations that the considerable amount of angular momentum (
inward or outward direction) would be transferred for various values of Gauss-Bonnet coupling
constant $\alpha$ either in negative or positive direction when it is compared with
the Kerr solution in the general relativity.

\begin{figure*}
 \vspace{1cm} 
  \center
  \psfig{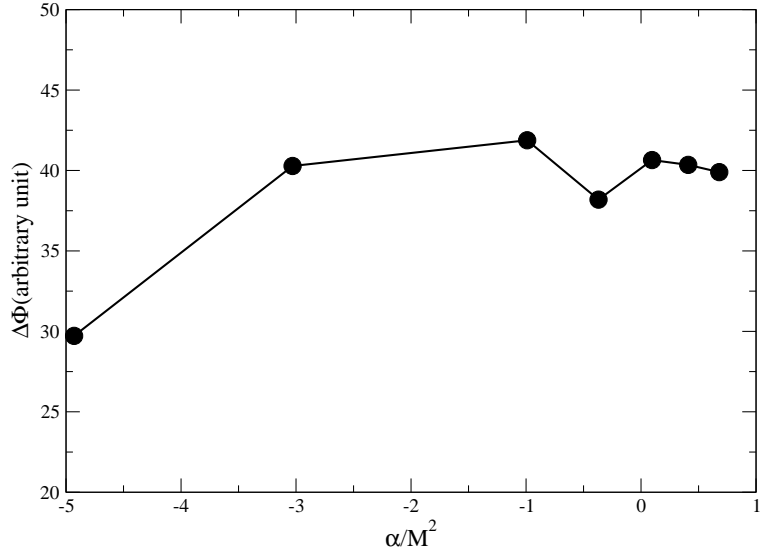}
  \caption{Dependencies of $\Delta \Phi$ versus Gauss-Bonnet coupling constant $\alpha$
    for a fixed black hole rotation parameter $a=0.28$ and at the radius $r_{in}=3.7M$.
    $\Delta \Phi$ gets maximum values on the left and the
    right sides  of $\alpha \rightarrow 0$ at where EGB black hole solution
    converges to the Kerr black hole one.}
\label{NR_5}
\end{figure*}

In order to depict  the strength of the chaotic behavior of the shock cone after reaching the steady
state, we plot $\Delta \Phi$ as a function of Gauss-Bonnet coupling constant $\alpha$
for a fixed black hole rotation parameter $a=0.28$ and  for the inner radius $r_{in}=3.7M$
in Fig. \ref{NR_5}. $\Delta \Phi  = \lvert A_{\max}-A_{min} \rvert$ represents  the
distance  between a crest and a through. It is obvious from the figure that the value of
$\Delta \Phi$ reaches the maximum  on the left and on the right sides  of $\alpha$
when $\alpha$ gets closer to  $0$. Moreover, the value of  $\Delta \Phi$ converges to Kerr solution
in general relativity when  $\alpha \rightarrow 0$.
It can also be seen in Table \ref{Inital_Con}. 

\begin{figure*}
 \vspace{1cm} 
  \center
  \psfig{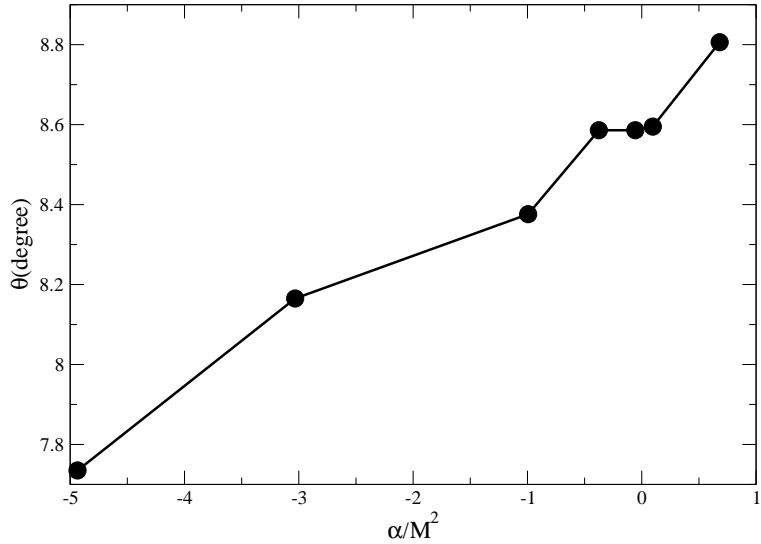}
  \caption{The dependency of the shock cone opening angle $\theta$ to the Gauss-Bonnet
    coupling constant $\alpha$ for the same initial parameters used in
    Fig.\ref{NR_5}. }
\label{NR_6}
\end{figure*}

In Fig.\ref{NR_6}, after the shock cone is accreated downstream region of the
accretion flow, the cone opening angle slightly depends on Gauss-Bonnet coupling constant $\alpha$
for the fixed value of the black hole rotation parameter $a=0.28$. While increasing  of $\alpha$
in the negative direction decreases the opening angle, this angle  slightly increases
with the increasing of $\alpha$ in the positive direction. It is also seen in Fig.\ref{NR_6} that
when $\alpha \rightarrow 0$, the black hole solution in EGB gravity converges to Kerr in
general relativity. It was also confirmed  for the Schwarzschild solution \citep{Donmez3}.
In particular,
as seen in Figs. \ref{NR_1}, \ref{NR_2}, and  \ref{NR_3},  the mass accretion rates strongly depend
on  $\alpha$ and $a$. In addition, the reduced shock opening angle would decrease in
the accretion rate \citep{Zanotti1}.

\begin{figure*}
 \vspace{1cm} 
  \center
  \psfig{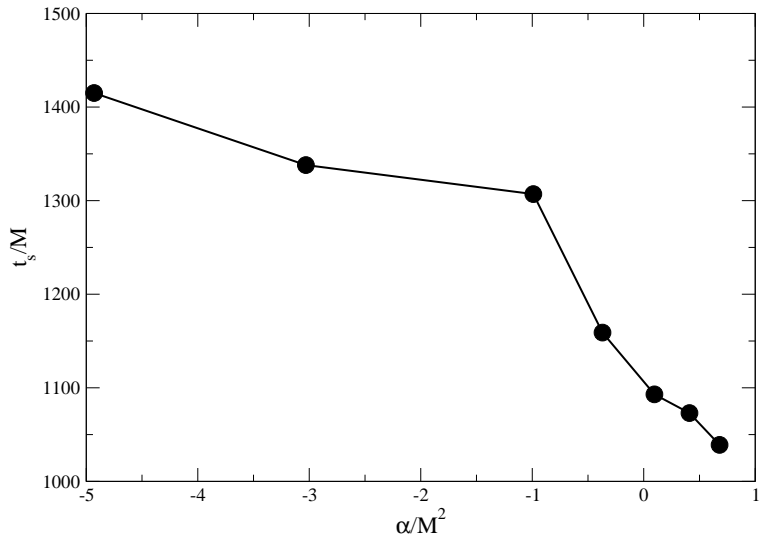}
  \caption{The time needed reach the quasi-steady-state (saturation time) versus
    the Gauss-Bonnet coupling constant $\alpha$ is plotted for the same
    initial parameters used in
    Fig.\ref{NR_5}. }
\label{NR_7}
\end{figure*}

In order to extract more information  about the availability of the shock cone
and its structure due to Gauss-Bonnet coupling constant $\alpha$,
we compute the required time  for the
shock cone  to go to the steady-state in Fig.\ref{NR_7}.
It is also called the saturation time.
According to our initial models, maximum saturation time is observed at
$\alpha \sim -5$ but it gets smaller when $\alpha$ is increasing. The same
behavior was also found for the non-rotating black hole in
EGB gravity \citep{Donmez3}. On the
other hand, it is seen in Table \ref{Inital_Con} that the saturation time
in EGB gravity converges to Kerr solution in general relativity.
In addition, the saturation time also depends on the black hole
rotation parameter $a$. The higher $a$ causes the lower the time to reach the
steady state, seen in Table \ref{Inital_Con}. 

\begin{figure*}
 \vspace{1cm} 
  \center
  \psfig{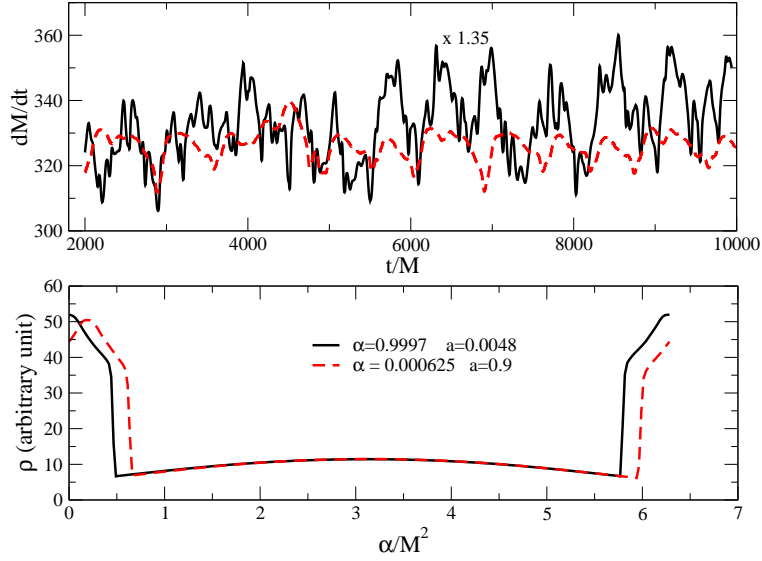}
  \caption{Comparisons of the mass accretion rates  after they have reached to the
    steady-state and the rest-mass densities along the angular direction at
    $r=\sim 6.5M$. We compare the numerical results between  the highest possible
    value of positive $\alpha$  with the smallest rotation parameter and the
    smallest possible value of positive $\alpha$ with a fastly rotating black hole.
     }
\label{NR_8}
\end{figure*}

In Fig.\ref{NR_8}, in order to illustrate the physical properties of shock cone as functions of  
Gauss-Bonnet coupling constant $\alpha$ and black hole rotation $a$, we consider
two different initial models. One is $\alpha=0.9997$ with very slowly
rotating black hole $a=0.0048$ and the other one is $\alpha=0.000625$ with a
fastly rotating black hole $a=0.9$.  We found that the shock cone location is shifted
to the right due to the warped space-time as a consequence of fastly rotating black hole,
seen in the bottom panel of Fig.\ref{NR_8}, when it is compared with the case, smaller rotation
parameter. In the upper panel of  Fig.\ref{NR_8}, one can observe that the
strong oscillation property can lead to more chaotic motions. These types of cases would
be a good candidate to observe QPOs in the $X-$ ray binary system. A similar behavior was also
found for the thin accretion disk around the non-rotating black hole in EGB gravity
\citep{Liu1}. They indicated that the disk is hotter and more efficient than that around
Schwarzschild black hole for a positive  $\alpha$. 
It is concluded that the shock cone location  is mainly dependent of black hole rotation
parameter $a$, while the wild behavior of the oscillating shock cone  is dependent of
Gauss-Bonnet coupling constant $\alpha$. Of course, the average values of mass accretion
rate for the fastly rotating black hole is $1.35$ times larger than the case  for very
slowly rotating one.

\subsection{Super-massive Black Hole $M87^*$}
\label{M87}

The existence of super-massive
black hole at the center of galaxy M87 has been observed using the very large
baseline interferometer \citep{Akiyama1, Akiyama2, Akiyama3, Akiyama4, Akiyama5}.
The observed $M87^*$ black hole shadow indicates that the observed data is well
consistent with the prediction of the general theory of relativity. The event
horizon telescope collaboration found the  black hole mass as
$M87^* = (6.5 \pm 0.7 )\times 10^9 M_{\odot}$. The calculated Schwarzschild radius of the
black hole is $R = 5.9 \times 10^{-4}$ parsecs.  The range of the radius can be defined
in terms of the black hole mass, $R = (1.899 \pm 0.206 ) M87^*$. The estimated rotation
parameter is $a=0.9 \pm 0.1$. It is consistent with the numerical simulations found by
using the numerical hydrodynamics with Kerr space-time metric
\citep{Akiyama1, Akiyama2, Akiyama3, Akiyama4, Akiyama5}. The black hole shadow
diameter was determined by the black hole mass-distance ratio and orientation of the
black hole spin axis. But many other effects were not considered which might also influence the 
the black hole diameter, such as the magnetic  filed  of the accretion flow \citep{Narayan1},
the electron heating and colliding processes \citep{Chael1, Davelaar1}, misalignment between
black hole spin and jet \citep{Vincent1}.

In addition to the effects of the general relativity,
the black hole shadow diameter could
be influenced by Gauss-Bonnet coupling constant $\alpha$ in EGB gravity.  Our numerical
results show that Gauss-Bonnet coupling constant could be used to constrain $M87^*$ radius for various
values of black hole rotation parameter. Constrained Gauss-Bonnet constant
should be $-1.7 <\alpha < \sim 0.35$ and the black bole rotation parameter
$a$ and black hole radius would vary depending on Gauss-Bonnet coupling constant as seen in
Table.\ref{Inital_Con}.  As a consequence, the possible size of black hole shadow
would be possible for varying  positive or negative $\alpha$ \citep{Minyong1}. The
negative Gauss-Bonnet constant $\alpha$ can break the universal bounds  on the
size of the black hole proposed in \citet{LuLyu1}.


\subsection{QPOs around the Black Hole in EGB gravity}
\label{QPOs around the Black Hole in EGB gravity}

The matter around the black hole is piled up due to the strong gravity and
causes strong collisions between gas molecules. So that the accreated material
is heated to high temperatures and QPOs would be created.  In the observed
$X-$ray fluxes, QPOs are commonly observed and they can be used to extract
the black hole properties, such as mass and spin indirectly.

\begin{figure*}
 \vspace{1cm} 
  \center
  \psfig{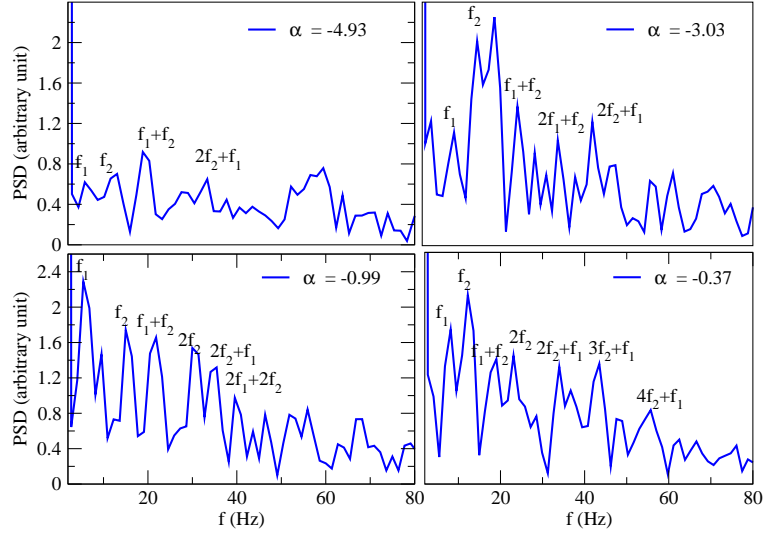}
  \caption{Power spectra is computed from the mass accretion rate dumped 
    at $r=6.5M$ by spherical detector for
    different values of Gauss-Bonnet coupling constant $\alpha$
    around the rotating black hole in EGB gravity with $a=0.28$. The
    black hole mass is chosen as $M = 10M_{\odot}$
     }
\label{QPOs1}
\end{figure*}

\begin{figure*}
 \vspace{1cm} 
  \center
  \psfig{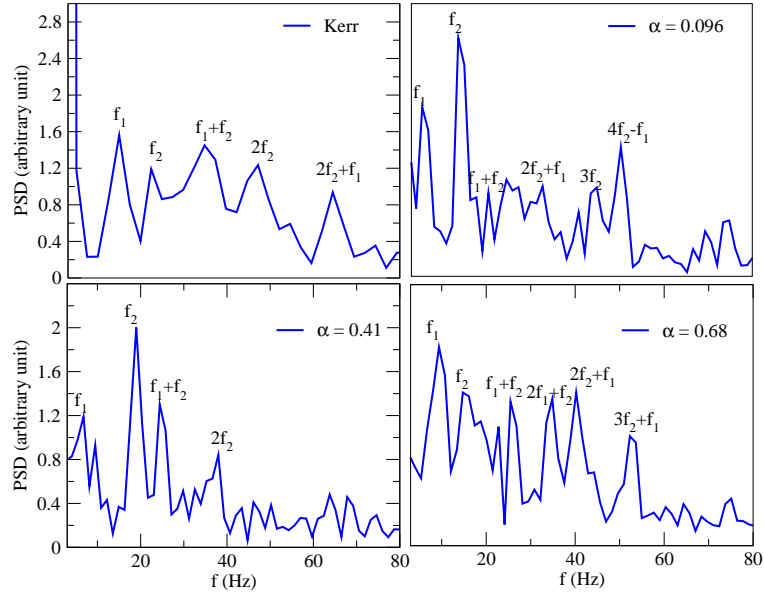}
  \caption{Same as Fig\ref{QPOs1} but for different values of $\alpha$ and
    Kerr solution in general relativity.
     }
\label{QPOs2}
\end{figure*}

The studying the QPOs in Fourier domain allows us to study the oscillation
properties of accreated matter and shock cone close to the black hole. So that
we can deduce the connections between oscillating shock and Gauss-Bonnet
coupling constant $\alpha$. The mass-accretion rate data is used to obtain the power
spectra  for different values of $\alpha$ and Kerr black hole, seen in
Figs.\ref{QPOs1} and \ref{QPOs2}. There is a distinct behavior
in the power spectra between Figs.\ref{QPOs1} and \ref{QPOs2}
computed inside the shock cone. The mode found in
the numerical solutions are global eigenmodes which occur in the oscillating shock cone
since the power spectra does not depend on the radial position. In Fig.\ref{QPOs1},
the genuine eigenmodes  and their nonlinear couplings are shown for the negative
values of $\alpha$, while these modes for  positive values of  $\alpha$ and Kerr black
hole are given in Fig.\ref{QPOs2}. For $\alpha = -4.93$, $f_1=6.1$ Hz and $f_2=12.9$ Hz
are genuine modes, while $f_1+f_2=19.7$ Hz and $f_1+2f_2=33.2$ Hz are the nonlinear
coupling of those genuine modes with a $1.2$ Hz of error bar. 
For $\alpha = -3.03$, $f_1=9.7$ Hz and $f_2=14.9$ Hz
are genuine modes, while $f_1+f_2=24.5$ Hz, $2f_1+f_2=34.26$ Hz, and $f_1+2f_2=42$ Hz
with a $2$ Hz of error bar.
For $\alpha = -0.99$, $f_1=5.5$ Hz and $f_2=15.2$ Hz
are genuine modes, while $f_1+f_2=21.6$ Hz, $2f_2=30.7$ Hz, $f_1+2f_2=35$ Hz, and $2f_1+2f_2=40$ Hz
with a $1$ Hz of error bar.
For $\alpha = -0.37$, $f_1=8.2$ Hz and $f_2=12.2$ Hz
are genuine modes, while $f_1+f_2=18.8$ Hz, $2f_2=23.2$ Hz, $f_1+2f_2=33.9$ Hz, $f_1+3f_2=43$ Hz,
and $f_1+4f_2=55.7$ Hz with a $2$ Hz of error bar.
For the shock cone around the Kerr black hole, $f_1=14.9$ Hz and $f_2=22.4$ Hz
are genuine modes, while $f_1+f_2=35.2$ Hz, $2f_2=47.2$ Hz, 
and $f_1+2f_2=64.4$ Hz with a $3$ Hz of error bar.
For $\alpha = 0.096$, $f_1=5.6$ Hz and $f_2=13.8$ Hz
are genuine modes, while $f_1+f_2=20.4$ Hz, $f_1+2f_2=32$ Hz, $3f_2=44$ Hz,
and $4f_2-f_1=50.4$ Hz with a $2$ Hz of error bar.
For $\alpha = 0.41$, $f_1=6.8$ Hz and $f_2=18.9$ Hz
are genuine modes, while $f_1+f_2=24.6$ Hz and  $2f_2=37.9$ Hz with a $1$ Hz of error bar.
For $\alpha = 0.68$, $f_1=9.3$ Hz and $f_2=15.3$ Hz
are genuine modes, while $f_1+f_2=25.7$ Hz, $f_1+2f_2=32$ Hz, $2f_1+f_2=34.8$ Hz ,
$f_1+2f_2=40.4$ Hz, and  $f_1+3f_2=52.7$ Hz with a $2$ Hz of error bar. The nonlinear
couplings of modes are the expected behavior in the nonlinear physical system
\citep{LL1976}. It is found that there is a correlation between Gauss-Bonnet coupling
constant $\alpha$ and genuine modes and their nonlinear coupling.  The amplitude of
the genuine mode is getting bigger when $\alpha$ approaches to zero in negative directions.
It was also confirmed in the oscillation of the shock cone around the non-rotating
black hole in EGB gravity \citet{Donmez3}. Similarly, the eigenmodes amplitudes are slightly
larger for bigger $\alpha$.  The influence of $\alpha$ on the power
eigenmodes and their nonlinear coupling is clearly seen in Figs.\ref{QPOs1} and \ref{QPOs2}.
The power spectrum density is more violent and shows more chaotic behavior for varying
values of $\alpha$ when we compare with Kerr black hole solution with the same rotation parameter
$a=0.28$. On the other hand, the nonlinear coupling term in the highest frequency is observed
in the Kerr solution.

\section{Conclusion}
\label{Conclusion}

We have performed a certain number of   numerical simulations of the shock
cone around the rotating black hole in EGB gravity by injecting the gas from
the upstream region of the computational domain  and by solving the GRH
equations using the HRSC scheme.
We have analyzed the dynamical structures of those cones and
their oscillation properties that can be affected by the
Gauss-Bonnet coupling constant $\alpha$ and black hole rotation parameter $a$.
The negative and positive values of $\alpha$ with the varying
$a$ are used to understand the many physical details of these shock cones.
The black hole solution in EGB gravity converges to Kerr in
general relativity when $\alpha \rightarrow 0$.

We find that the  more matter close the black hole is not
only piled up due to the fastly rotating black hole but the different
values of $\alpha$ also causes to it. Increasing in the amount of  matter
lades a chaotic motion and also causes the matter falling 
towards the black hole.
After the shock cone reaches to the steady state, 
the bigger mass accretion rate gradient would lead more chaotic
radiation in the observed phenomena. 
On the other hand, the position of shock cone
is mainly dependent of the black hole rotation
parameter $a$, while the wild behavior of the oscillating shock cone  is 
dependent of Gauss-Bonnet coupling constant $\alpha$. 
Increasing  $\alpha$
in the negative direction decreases the opening angle
of the shock cone, this angle  slightly increases
with the increasing  $\alpha$ in the positive direction.

Together with the mass accretion rate towards the black hole, we have
investigated the angular momentum rate, which is particularly important to
know whether  the gas moves towards or away from the black hole. Irregularities
would be observed on the shock cone when the angular momentum transformation occurs.
Our simulations show  that the considerable amount of angular momentum (
inward or outward direction) would be transferred for various
values of Gauss-Bonnet coupling
constant $\alpha$ either in negative or positive direction when it is compared with
the Kerr solution in the general relativity. 

We have also focused on the oscillations inside the shock cones after
it reaches the steady state. The saturation times to reach  the steady
state are not the same in all models.  The maximum saturation time is observed at
$\alpha \sim -5$ and it gets smaller when $\alpha$ increases. On the
other hand, the saturation time
in EGB gravity converges to Kerr solution in general relativity.
Meanwhile, the oscillation properties of the accreated matter and shock
cone close to the black hole can be extracted
using the Fourier transform to compute the power spectrum density .
The spectrum density shows the non-linear couplings of the modes for varying
values of $\alpha$ when we compare with Kerr black hole solution with the same
rotation parameter and black hole mass.

Finally, the alternative theory of the modified gravity should not be ignored
to constrain the physical properties of the observed black hole. 
We have carried out an indirect comparison with observed shadow of $M87^*$
black hole and found that Gauss-Bonnet coupling constant $\alpha$
could be used to constrain $M87^*$ radius for various  values of the black
hole rotation parameter $a$. Our numerical simulations show that 
the smaller  $a$ and a bigger shadow radius could be used 
to understand the physical properties of the $M87^*$.


\section*{Acknowledgments}
All simulations were performed using the Phoenix  High
Performance Computing facility at the American University of the Middle East
(AUM), Kuwait.\\

\bibliography{reference.bib}

\begin{thebibliography}{}
\expandafter\ifx\csname natexlab\endcsname\relax\def\natexlab#1{#1}\fi

\bibitem[{{Azreg-A{\"\i}nou}(2014)}]{Azreg1}
{Azreg-A{\"\i}nou}, M. 2014, \prd, 90, 064041

\bibitem[{{Bambi} {et~al.}(2019){Bambi}, {Freese}, {Vagnozzi}, \&
  {Visinelli}}]{Bambi1}
{Bambi}, C., {Freese}, K., {Vagnozzi}, S., \& {Visinelli}, L. 2019, \prd, 100,
  044057

\bibitem[{Blondin(2013)}]{Blondin1}
Blondin, J.~M. 2013, The Astrophysical Journal, 767, 135

\bibitem[{{Bondi} \& {Hoyle}(1944)}]{Bondi1}
{Bondi}, H., \& {Hoyle}, F. 1944, \mnras, 104, 273

\bibitem[{{Chael} {et~al.}(2018){Chael}, {Rowan}, {Narayan}, {Johnson}, \&
  {Sironi}}]{Chael1}
{Chael}, A., {Rowan}, M., {Narayan}, R., {Johnson}, M., \& {Sironi}, L. 2018,
  \mnras, 478, 5209

\bibitem[{{Clifton} {et~al.}(2020){Clifton}, {Carrilho}, {Fernandes}, \&
  {Mulryne}}]{Timothy1}
{Clifton}, T., {Carrilho}, P., {Fernandes}, P. G.~S., \& {Mulryne}, D.~J. 2020,
  \prd, 102, 084005

\bibitem[{{Cruz-Osorio} \& {Lora-Clavijo}(2016)}]{CruzOsorio1}
{Cruz-Osorio}, A., \& {Lora-Clavijo}, F.~D. 2016, \mnras, 460, 3193

\bibitem[{{Davelaar} {et~al.}(2019){Davelaar}, {Olivares}, {Porth},
  {Bronzwaer}, {Janssen}, {Roelofs}, {Mizuno}, {Fromm}, {Falcke}, \&
  {Rezzolla}}]{Davelaar1}
{Davelaar}, J., {Olivares}, H., {Porth}, O., {et~al.} 2019, \aap, 632, A2

\bibitem[{Davies \& Pringle(1980)}]{Davies1}
Davies, R.~E., \& Pringle, J.~E. 1980, Monthly Notices of the Royal
  Astronomical Society, 191, 599

\bibitem[{{D{\"o}nmez}(2004)}]{Donmez1}
{D{\"o}nmez}, O. 2004, \apss, 293, 323

\bibitem[{{Donmez}(2006)}]{Donmez2}
{Donmez}, O. 2006, AM\&C, 181, 256

\bibitem[{{D{\"o}nmez}(2012)}]{Donmez5}
{D{\"o}nmez}, O. 2012, \mnras, 426, 1533

\bibitem[{{D{\"o}nmez} {et~al.}(2011){D{\"o}nmez}, {Zanotti}, \&
  {Rezzolla}}]{Donmez6}
{D{\"o}nmez}, O., {Zanotti}, O., \& {Rezzolla}, L. 2011, \mnras, 412, 1659

\bibitem[{{Donmez, Orhan}(2021)}]{Donmez3}
{Donmez, Orhan}. 2021, Eur. Phys. J. C, 81, 113

\bibitem[{{Event Horizon Telescope Collaboration}
  {et~al.}(2019{\natexlab{a}}){Event Horizon Telescope Collaboration},
  {Akiyama}, {Alberdi}, {Alef}, {Asada}, {Azulay}, {Baczko}, {Ball},
  {Balokovi{\'c}}, {Barrett}, {Bintley}, {Blackburn}, {Boland}, {Bouman},
  {Bower}, {Bremer}, {Brinkerink}, {Brissenden}, {Britzen}, {Broderick},
  {Broguiere}, {Bronzwaer}, {Byun}, {Carlstrom}, {Chael}, {Chan}, {Chatterjee},
  {Chatterjee}, {Chen}, {Chen}, {Cho}, {Christian}, {Conway}, {Cordes}, {Crew},
  {Cui}, {Davelaar}, {De Laurentis}, {Deane}, {Dempsey}, {Desvignes}, {Dexter},
  {Doeleman}, {Eatough}, {Falcke}, {Fish}, {Fomalont}, {Fraga-Encinas},
  {Freeman}, {Friberg}, {Fromm}, {G{\'o}mez}, {Galison}, {Gammie},
  {Garc{\'\i}a}, {Gentaz}, {Georgiev}, {Goddi}, {Gold}, {Gu}, {Gurwell},
  {Hada}, {Hecht}, {Hesper}, {Ho}, {Ho}, {Honma}, {Huang}, {Huang}, {Hughes},
  {Ikeda}, {Inoue}, {Issaoun}, {James}, {Jannuzi}, {Janssen}, {Jeter}, {Jiang},
  {Johnson}, {Jorstad}, {Jung}, {Karami}, {Karuppusamy}, {Kawashima},
  {Keating}, {Kettenis}, {Kim}, {Kim}, {Kim}, {Kino}, {Koay}, {Koch}, {Koyama},
  {Kramer}, {Kramer}, {Krichbaum}, {Kuo}, {Lauer}, {Lee}, {Li}, {Li},
  {Lindqvist}, {Liu}, {Liuzzo}, {Lo}, {Lobanov}, {Loinard}, {Lonsdale}, {Lu},
  {MacDonald}, {Mao}, {Markoff}, {Marrone}, {Marscher}, {Mart{\'\i}-Vidal},
  {Matsushita}, {Matthews}, {Medeiros}, {Menten}, {Mizuno}, {Mizuno}, {Moran},
  {Moriyama}, {Moscibrodzka}, {M{\"u}ller}, {Nagai}, {Nagar}, {Nakamura},
  {Narayan}, {Narayanan}, {Natarajan}, {Neri}, {Ni}, {Noutsos}, {Okino},
  {Olivares}, {Ortiz-Le{\'o}n}, {Oyama}, {{\"O}zel}, {Palumbo}, {Patel}, {Pen},
  {Pesce}, {Pi{\'e}tu}, {Plambeck}, {PopStefanija}, {Porth}, {Prather},
  {Preciado-L{\'o}pez}, {Psaltis}, {Pu}, {Ramakrishnan}, {Rao}, {Rawlings},
  {Raymond}, {Rezzolla}, {Ripperda}, {Roelofs}, {Rogers}, {Ros}, {Rose},
  {Roshanineshat}, {Rottmann}, {Roy}, {Ruszczyk}, {Ryan}, {Rygl},
  {S{\'a}nchez}, {S{\'a}nchez-Arguelles}, {Sasada}, {Savolainen}, {Schloerb},
  {Schuster}, {Shao}, {Shen}, {Small}, {Sohn}, {SooHoo}, {Tazaki}, {Tiede},
  {Tilanus}, {Titus}, {Toma}, {Torne}, {Trent}, {Trippe}, {Tsuda}, {van
  Bemmel}, {van Langevelde}, {van Rossum}, {Wagner}, {Wardle}, {Weintroub},
  {Wex}, {Wharton}, {Wielgus}, {Wong}, {Wu}, {Young}, {Young}, {Younsi},
  {Yuan}, {Yuan}, {Zensus}, {Zhao}, {Zhao}, {Zhu}, {Algaba}, {Allardi},
  {Amestica}, {Anczarski}, {Bach}, {Baganoff}, {Beaudoin}, {Benson},
  {Berthold}, {Blanchard}, {Blundell}, {Bustamente}, {Cappallo},
  {Castillo-Dom{\'\i}nguez}, {Chang}, {Chang}, {Chang}, {Chen}, {Chilson},
  {Chuter}, {C{\'o}rdova Rosado}, {Coulson}, {Crawford}, {Crowley}, {David},
  {Derome}, {Dexter}, {Dornbusch}, {Dudevoir}, {Dzib}, {Eckart}, {Eckert},
  {Erickson}, {Everett}, {Faber}, {Farah}, {Fath}, {Folkers}, {Forbes},
  {Freund}, {G{\'o}mez-Ruiz}, {Gale}, {Gao}, {Geertsema}, {Graham}, {Greer},
  {Grosslein}, {Gueth}, {Haggard}, {Halverson}, {Han}, {Han}, {Hao},
  {Hasegawa}, {Henning}, {Hern{\'a}ndez-G{\'o}mez}, {Herrero-Illana},
  {Heyminck}, {Hirota}, {Hoge}, {Huang}, {Impellizzeri}, {Jiang}, {Kamble},
  {Keisler}, {Kimura}, {Kono}, {Kubo}, {Kuroda}, {Lacasse}, {Laing}, {Leitch},
  {Li}, {Lin}, {Liu}, {Liu}, {Lu}, {Marson}, {Martin-Cocher}, {Massingill},
  {Matulonis}, {McColl}, {McWhirter}, {Messias}, {Meyer-Zhao}, {Michalik},
  {Monta{\~n}a}, {Montgomerie}, {Mora-Klein}, {Muders}, {Nadolski}, {Navarro},
  {Neilsen}, {Nguyen}, {Nishioka}, {Norton}, {Nowak}, {Nystrom}, {Ogawa},
  {Oshiro}, {Oyama}, {Parsons}, {Paine}, {Pe{\~n}alver}, {Phillips}, {Poirier},
  {Pradel}, {Primiani}, {Raffin}, {Rahlin}, {Reiland}, {Risacher}, {Ruiz},
  {S{\'a}ez-Mada{\'\i}n}, {Sassella}, {Schellart}, {Shaw}, {Silva}, {Shiokawa},
  {Smith}, {Snow}, {Souccar}, {Sousa}, {Sridharan}, {Srinivasan}, {Stahm},
  {Stark}, {Story}, {Timmer}, {Vertatschitsch}, {Walther}, {Wei}, {Whitehorn},
  {Whitney}, {Woody}, {Wouterloot}, {Wright}, {Yamaguchi}, {Yu}, {Zeballos},
  {Zhang}, \& {Ziurys}}]{Akiyama1}
{Event Horizon Telescope Collaboration}, {Akiyama}, K., {Alberdi}, A., {et~al.}
  2019{\natexlab{a}}, \apjl, 875, L1

\bibitem[{{Event Horizon Telescope Collaboration}
  {et~al.}(2019{\natexlab{b}}){Event Horizon Telescope Collaboration},
  {Akiyama}, {Alberdi}, {Alef}, {Asada}, {Azulay}, {Baczko}, {Ball},
  {Balokovi{\'c}}, {Barrett}, {Bintley}, {Blackburn}, {Boland}, {Bouman},
  {Bower}, {Bremer}, {Brinkerink}, {Brissenden}, {Britzen}, {Broderick},
  {Broguiere}, {Bronzwaer}, {Byun}, {Carlstrom}, {Chael}, {Chan}, {Chatterjee},
  {Chatterjee}, {Chen}, {Chen}, {Cho}, {Christian}, {Conway}, {Cordes}, {Crew},
  {Cui}, {Davelaar}, {De Laurentis}, {Deane}, {Dempsey}, {Desvignes}, {Dexter},
  {Doeleman}, {Eatough}, {Falcke}, {Fish}, {Fomalont}, {Fraga-Encinas},
  {Friberg}, {Fromm}, {G{\'o}mez}, {Galison}, {Gammie}, {Garc{\'\i}a},
  {Gentaz}, {Georgiev}, {Goddi}, {Gold}, {Gu}, {Gurwell}, {Hada}, {Hecht},
  {Hesper}, {Ho}, {Ho}, {Honma}, {Huang}, {Huang}, {Hughes}, {Ikeda}, {Inoue},
  {Issaoun}, {James}, {Jannuzi}, {Janssen}, {Jeter}, {Jiang}, {Johnson},
  {Jorstad}, {Jung}, {Karami}, {Karuppusamy}, {Kawashima}, {Keating},
  {Kettenis}, {Kim}, {Kim}, {Kim}, {Kino}, {Koay}, {Koch}, {Koyama}, {Kramer},
  {Kramer}, {Krichbaum}, {Kuo}, {Lauer}, {Lee}, {Li}, {Li}, {Lindqvist}, {Liu},
  {Liuzzo}, {Lo}, {Lobanov}, {Loinard}, {Lonsdale}, {Lu}, {MacDonald}, {Mao},
  {Markoff}, {Marrone}, {Marscher}, {Mart{\'\i}-Vidal}, {Matsushita},
  {Matthews}, {Medeiros}, {Menten}, {Mizuno}, {Mizuno}, {Moran}, {Moriyama},
  {Moscibrodzka}, {M{\"u}ller}, {Nagai}, {Nagar}, {Nakamura}, {Narayan},
  {Narayanan}, {Natarajan}, {Neri}, {Ni}, {Noutsos}, {Okino}, {Olivares},
  {Ortiz-Le{\'o}n}, {Oyama}, {{\"O}zel}, {Palumbo}, {Patel}, {Pen}, {Pesce},
  {Pi{\'e}tu}, {Plambeck}, {PopStefanija}, {Porth}, {Prather},
  {Preciado-L{\'o}pez}, {Psaltis}, {Pu}, {Ramakrishnan}, {Rao}, {Rawlings},
  {Raymond}, {Rezzolla}, {Ripperda}, {Roelofs}, {Rogers}, {Ros}, {Rose},
  {Roshanineshat}, {Rottmann}, {Roy}, {Ruszczyk}, {Ryan}, {Rygl},
  {S{\'a}nchez}, {S{\'a}nchez-Arguelles}, {Sasada}, {Savolainen}, {Schloerb},
  {Schuster}, {Shao}, {Shen}, {Small}, {Sohn}, {SooHoo}, {Tazaki}, {Tiede},
  {Tilanus}, {Titus}, {Toma}, {Torne}, {Trent}, {Trippe}, {Tsuda}, {van
  Bemmel}, {van Langevelde}, {van Rossum}, {Wagner}, {Wardle}, {Weintroub},
  {Wex}, {Wharton}, {Wielgus}, {Wong}, {Wu}, {Young}, {Young}, {Younsi},
  {Yuan}, {Yuan}, {Zensus}, {Zhao}, {Zhao}, {Zhu}, {Algaba}, {Allardi},
  {Amestica}, {Bach}, {Beaudoin}, {Benson}, {Berthold}, {Blanchard},
  {Blundell}, {Bustamente}, {Cappallo}, {Castillo-Dom{\'\i}nguez}, {Chang},
  {Chang}, {Chang}, {Chen}, {Chilson}, {Chuter}, {C{\'o}rdova Rosado},
  {Coulson}, {Crawford}, {Crowley}, {David}, {Derome}, {Dexter}, {Dornbusch},
  {Dudevoir}, {Dzib}, {Eckert}, {Erickson}, {Everett}, {Faber}, {Farah},
  {Fath}, {Folkers}, {Forbes}, {Freund}, {G{\'o}mez-Ruiz}, {Gale}, {Gao},
  {Geertsema}, {Graham}, {Greer}, {Grosslein}, {Gueth}, {Halverson}, {Han},
  {Han}, {Hao}, {Hasegawa}, {Henning}, {Hern{\'a}ndez-G{\'o}mez},
  {Herrero-Illana}, {Heyminck}, {Hirota}, {Hoge}, {Huang}, {Impellizzeri},
  {Jiang}, {Kamble}, {Keisler}, {Kimura}, {Kono}, {Kubo}, {Kuroda}, {Lacasse},
  {Laing}, {Leitch}, {Li}, {Lin}, {Liu}, {Liu}, {Lu}, {Marson},
  {Martin-Cocher}, {Massingill}, {Matulonis}, {McColl}, {McWhirter}, {Messias},
  {Meyer-Zhao}, {Michalik}, {Monta{\~n}a}, {Montgomerie}, {Mora-Klein},
  {Muders}, {Nadolski}, {Navarro}, {Nguyen}, {Nishioka}, {Norton}, {Nystrom},
  {Ogawa}, {Oshiro}, {Oyama}, {Padin}, {Parsons}, {Paine}, {Pe{\~n}alver},
  {Phillips}, {Poirier}, {Pradel}, {Primiani}, {Raffin}, {Rahlin}, {Reiland},
  {Risacher}, {Ruiz}, {S{\'a}ez-Mada{\'\i}n}, {Sassella}, {Schellart}, {Shaw},
  {Silva}, {Shiokawa}, {Smith}, {Snow}, {Souccar}, {Sousa}, {Sridharan},
  {Srinivasan}, {Stahm}, {Stark}, {Story}, {Timmer}, {Vertatschitsch},
  {Walther}, {Wei}, {Whitehorn}, {Whitney}, {Woody}, {Wouterloot}, {Wright},
  {Yamaguchi}, {Yu}, {Zeballos}, \& {Ziurys}}]{Akiyama2}
---. 2019{\natexlab{b}}, \apjl, 875, L2

\bibitem[{{Event Horizon Telescope Collaboration}
  {et~al.}(2019{\natexlab{c}}){Event Horizon Telescope Collaboration},
  {Akiyama}, {Alberdi}, {Alef}, {Asada}, {Azulay}, {Baczko}, {Ball},
  {Balokovi{\'c}}, {Barrett}, {Bintley}, {Blackburn}, {Boland}, {Bouman},
  {Bower}, {Bremer}, {Brinkerink}, {Brissenden}, {Britzen}, {Broderick},
  {Broguiere}, {Bronzwaer}, {Byun}, {Carlstrom}, {Chael}, {Chan}, {Chatterjee},
  {Chatterjee}, {Chen}, {Chen}, {Cho}, {Christian}, {Conway}, {Cordes}, {Crew},
  {Cui}, {Davelaar}, {De Laurentis}, {Deane}, {Dempsey}, {Desvignes}, {Dexter},
  {Doeleman}, {Eatough}, {Falcke}, {Fish}, {Fomalont}, {Fraga-Encinas},
  {Friberg}, {Fromm}, {G{\'o}mez}, {Galison}, {Gammie}, {Garc{\'\i}a},
  {Gentaz}, {Georgiev}, {Goddi}, {Gold}, {Gu}, {Gurwell}, {Hada}, {Hecht},
  {Hesper}, {Ho}, {Ho}, {Honma}, {Huang}, {Huang}, {Hughes}, {Ikeda}, {Inoue},
  {Issaoun}, {James}, {Jannuzi}, {Janssen}, {Jeter}, {Jiang}, {Johnson},
  {Jorstad}, {Jung}, {Karami}, {Karuppusamy}, {Kawashima}, {Keating},
  {Kettenis}, {Kim}, {Kim}, {Kim}, {Kino}, {Koay}, {Koch}, {Koyama}, {Kramer},
  {Kramer}, {Krichbaum}, {Kuo}, {Lauer}, {Lee}, {Li}, {Li}, {Lindqvist}, {Liu},
  {Liuzzo}, {Lo}, {Lobanov}, {Loinard}, {Lonsdale}, {Lu}, {MacDonald}, {Mao},
  {Markoff}, {Marrone}, {Marscher}, {Mart{\'\i}-Vidal}, {Matsushita},
  {Matthews}, {Medeiros}, {Menten}, {Mizuno}, {Mizuno}, {Moran}, {Moriyama},
  {Moscibrodzka}, {M{\"u}ller}, {Nagai}, {Nagar}, {Nakamura}, {Narayan},
  {Narayanan}, {Natarajan}, {Neri}, {Ni}, {Noutsos}, {Okino}, {Olivares},
  {Ortiz-Le{\'o}n}, {Oyama}, {{\"O}zel}, {Palumbo}, {Patel}, {Pen}, {Pesce},
  {Pi{\'e}tu}, {Plambeck}, {PopStefanija}, {Porth}, {Prather},
  {Preciado-L{\'o}pez}, {Psaltis}, {Pu}, {Ramakrishnan}, {Rao}, {Rawlings},
  {Raymond}, {Rezzolla}, {Ripperda}, {Roelofs}, {Rogers}, {Ros}, {Rose},
  {Roshanineshat}, {Rottmann}, {Roy}, {Ruszczyk}, {Ryan}, {Rygl},
  {S{\'a}nchez}, {S{\'a}nchez-Arguelles}, {Sasada}, {Savolainen}, {Schloerb},
  {Schuster}, {Shao}, {Shen}, {Small}, {Sohn}, {SooHoo}, {Tazaki}, {Tiede},
  {Tilanus}, {Titus}, {Toma}, {Torne}, {Trent}, {Trippe}, {Tsuda}, {van
  Bemmel}, {van Langevelde}, {van Rossum}, {Wagner}, {Wardle}, {Weintroub},
  {Wex}, {Wharton}, {Wielgus}, {Wong}, {Wu}, {Young}, {Young}, {Younsi},
  {Yuan}, {Yuan}, {Zensus}, {Zhao}, {Zhao}, {Zhu}, {Cappallo}, {Farah},
  {Folkers}, {Meyer-Zhao}, {Michalik}, {Nadolski}, {Nishioka}, {Pradel},
  {Primiani}, {Souccar}, {Vertatschitsch}, \& {Yamaguchi}}]{Akiyama3}
---. 2019{\natexlab{c}}, \apjl, 875, L3

\bibitem[{{Event Horizon Telescope Collaboration}
  {et~al.}(2019{\natexlab{d}}){Event Horizon Telescope Collaboration},
  {Akiyama}, {Alberdi}, {Alef}, {Asada}, {Azulay}, {Baczko}, {Ball},
  {Balokovi{\'c}}, {Barrett}, {Bintley}, {Blackburn}, {Boland}, {Bouman},
  {Bower}, {Bremer}, {Brinkerink}, {Brissenden}, {Britzen}, {Broderick},
  {Broguiere}, {Bronzwaer}, {Byun}, {Carlstrom}, {Chael}, {Chan}, {Chatterjee},
  {Chatterjee}, {Chen}, {Chen}, {Cho}, {Christian}, {Conway}, {Cordes}, {Crew},
  {Cui}, {Davelaar}, {De Laurentis}, {Deane}, {Dempsey}, {Desvignes}, {Dexter},
  {Doeleman}, {Eatough}, {Falcke}, {Fish}, {Fomalont}, {Fraga-Encinas},
  {Freeman}, {Friberg}, {Fromm}, {G{\'o}mez}, {Galison}, {Gammie},
  {Garc{\'\i}a}, {Gentaz}, {Georgiev}, {Goddi}, {Gold}, {Gu}, {Gurwell},
  {Hada}, {Hecht}, {Hesper}, {Ho}, {Ho}, {Honma}, {Huang}, {Huang}, {Hughes},
  {Ikeda}, {Inoue}, {Issaoun}, {James}, {Jannuzi}, {Janssen}, {Jeter}, {Jiang},
  {Johnson}, {Jorstad}, {Jung}, {Karami}, {Karuppusamy}, {Kawashima},
  {Keating}, {Kettenis}, {Kim}, {Kim}, {Kim}, {Kino}, {Koay}, {Koch}, {Koyama},
  {Kramer}, {Kramer}, {Krichbaum}, {Kuo}, {Lauer}, {Lee}, {Li}, {Li},
  {Lindqvist}, {Liu}, {Liuzzo}, {Lo}, {Lobanov}, {Loinard}, {Lonsdale}, {Lu},
  {MacDonald}, {Mao}, {Markoff}, {Marrone}, {Marscher}, {Mart{\'\i}-Vidal},
  {Matsushita}, {Matthews}, {Medeiros}, {Menten}, {Mizuno}, {Mizuno}, {Moran},
  {Moriyama}, {Moscibrodzka}, {M{\"u}ller}, {Nagai}, {Nagar}, {Nakamura},
  {Narayan}, {Narayanan}, {Natarajan}, {Neri}, {Ni}, {Noutsos}, {Okino},
  {Olivares}, {Oyama}, {{\"O}zel}, {Palumbo}, {Patel}, {Pen}, {Pesce},
  {Pi{\'e}tu}, {Plambeck}, {PopStefanija}, {Porth}, {Prather},
  {Preciado-L{\'o}pez}, {Psaltis}, {Pu}, {Ramakrishnan}, {Rao}, {Rawlings},
  {Raymond}, {Rezzolla}, {Ripperda}, {Roelofs}, {Rogers}, {Ros}, {Rose},
  {Roshanineshat}, {Rottmann}, {Roy}, {Ruszczyk}, {Ryan}, {Rygl},
  {S{\'a}nchez}, {S{\'a}nchez-Arguelles}, {Sasada}, {Savolainen}, {Schloerb},
  {Schuster}, {Shao}, {Shen}, {Small}, {Sohn}, {SooHoo}, {Tazaki}, {Tiede},
  {Tilanus}, {Titus}, {Toma}, {Torne}, {Trent}, {Trippe}, {Tsuda}, {van
  Bemmel}, {van Langevelde}, {van Rossum}, {Wagner}, {Wardle}, {Weintroub},
  {Wex}, {Wharton}, {Wielgus}, {Wong}, {Wu}, {Young}, {Young}, {Younsi},
  {Yuan}, {Yuan}, {Zensus}, {Zhao}, {Zhao}, {Zhu}, {Farah}, {Meyer-Zhao},
  {Michalik}, {Nadolski}, {Nishioka}, {Pradel}, {Primiani}, {Souccar},
  {Vertatschitsch}, \& {Yamaguchi}}]{Akiyama4}
---. 2019{\natexlab{d}}, \apjl, 875, L4

\bibitem[{{Event Horizon Telescope Collaboration}
  {et~al.}(2019{\natexlab{e}}){Event Horizon Telescope Collaboration},
  {Akiyama}, {Alberdi}, {Alef}, {Asada}, {Azulay}, {Baczko}, {Ball},
  {Balokovi{\'c}}, {Barrett}, {Bintley}, {Blackburn}, {Boland}, {Bouman},
  {Bower}, {Bremer}, {Brinkerink}, {Brissenden}, {Britzen}, {Broderick},
  {Broguiere}, {Bronzwaer}, {Byun}, {Carlstrom}, {Chael}, {Chan}, {Chatterjee},
  {Chatterjee}, {Chen}, {Chen}, {Cho}, {Christian}, {Conway}, {Cordes}, {Crew},
  {Cui}, {Davelaar}, {De Laurentis}, {Deane}, {Dempsey}, {Desvignes}, {Dexter},
  {Doeleman}, {Eatough}, {Falcke}, {Fish}, {Fomalont}, {Fraga-Encinas},
  {Friberg}, {Fromm}, {G{\'o}mez}, {Galison}, {Gammie}, {Garc{\'\i}a},
  {Gentaz}, {Georgiev}, {Goddi}, {Gold}, {Gu}, {Gurwell}, {Hada}, {Hecht},
  {Hesper}, {Ho}, {Ho}, {Honma}, {Huang}, {Huang}, {Hughes}, {Ikeda}, {Inoue},
  {Issaoun}, {James}, {Jannuzi}, {Janssen}, {Jeter}, {Jiang}, {Johnson},
  {Jorstad}, {Jung}, {Karami}, {Karuppusamy}, {Kawashima}, {Keating},
  {Kettenis}, {Kim}, {Kim}, {Kim}, {Kino}, {Koay}, {Koch}, {Koyama}, {Kramer},
  {Kramer}, {Krichbaum}, {Kuo}, {Lauer}, {Lee}, {Li}, {Li}, {Lindqvist}, {Liu},
  {Liuzzo}, {Lo}, {Lobanov}, {Loinard}, {Lonsdale}, {Lu}, {MacDonald}, {Mao},
  {Markoff}, {Marrone}, {Marscher}, {Mart{\'\i}-Vidal}, {Matsushita},
  {Matthews}, {Medeiros}, {Menten}, {Mizuno}, {Mizuno}, {Moran}, {Moriyama},
  {Moscibrodzka}, {Mul{\ensuremath{\ddot{}}}ler}, {Nagai}, {Nagar}, {Nakamura},
  {Narayan}, {Narayanan}, {Natarajan}, {Neri}, {Ni}, {Noutsos}, {Okino},
  {Olivares}, {Oyama}, {{\"O}zel}, {Palumbo}, {Patel}, {Pen}, {Pesce},
  {Pi{\'e}tu}, {Plambeck}, {PopStefanija}, {Porth}, {Prather},
  {Preciado-L{\'o}pez}, {Psaltis}, {Pu}, {Ramakrishnan}, {Rao}, {Rawlings},
  {Raymond}, {Rezzolla}, {Ripperda}, {Roelofs}, {Rogers}, {Ros}, {Rose},
  {Roshanineshat}, {Rottmann}, {Roy}, {Ruszczyk}, {Ryan}, {Rygl},
  {S{\'a}nchez}, {S{\'a}nchez-Arguelles}, {Sasada}, {Savolainen}, {Schloerb},
  {Schuster}, {Shao}, {Shen}, {Small}, {Sohn}, {SooHoo}, {Tazaki}, {Tiede},
  {Tilanus}, {Titus}, {Toma}, {Torne}, {Trent}, {Trippe}, {Tsuda}, {van
  Bemmel}, {van Langevelde}, {van Rossum}, {Wagner}, {Wardle}, {Weintroub},
  {Wex}, {Wharton}, {Wielgus}, {Wong}, {Wu}, {Young}, {Young}, {Younsi},
  {Yuan}, {Yuan}, {Zensus}, {Zhao}, {Zhao}, {Zhu}, {Anczarski}, {Baganoff},
  {Eckart}, {Farah}, {Haggard}, {Meyer-Zhao}, {Michalik}, {Nadolski},
  {Neilsen}, {Nishioka}, {Nowak}, {Pradel}, {Primiani}, {Souccar},
  {Vertatschitsch}, {Yamaguchi}, \& {Zhang}}]{Akiyama5}
---. 2019{\natexlab{e}}, \apjl, 875, L5

\bibitem[{{Feng} {et~al.}(2020){Feng}, {Gu}, \& {Shu}}]{Feng1}
{Feng}, J.-X., {Gu}, B.-M., \& {Shu}, F.-W. 2020, arXiv e-prints,
  arXiv:2006.16751

\bibitem[{{Foglizzo} {et~al.}(2005){Foglizzo}, {Galletti}, \&
  {Ruffert}}]{Foglizzo1}
{Foglizzo}, T., {Galletti}, P., \& {Ruffert}, M. 2005, \aap, 435, 397

\bibitem[{{Ghosh} {et~al.}(2020){Ghosh}, {Kumar}, \& {Singh}}]{Ghosh1}
{Ghosh}, S.~G., {Kumar}, A., \& {Singh}, D.~V. 2020, Physics of the Dark
  Universe, 30, 100660

\bibitem[{{Ghosh} \& {Kumar}(2020)}]{Ghosh2}
{Ghosh}, S.~G., \& {Kumar}, R. 2020, Classical and Quantum Gravity, 37, 245008

\bibitem[{Ghosh \& Maharaj(2020)}]{Sushant1}
Ghosh, S.~G., \& Maharaj, S.~D. 2020, Physics of the Dark Universe, 30, 100687

\bibitem[{{Glavan} \& {Lin}(2020)}]{Glavan1}
{Glavan}, D., \& {Lin}, C. 2020, \prl, 124, 081301

\bibitem[{{Guo} \& {Li}(2020)}]{Minyong1}
{Guo}, M., \& {Li}, P.-C. 2020, European Physical Journal C, 80, 588

\bibitem[{Haghani(2020)}]{Zahra1}
Haghani, Z. 2020, Physics of the Dark Universe, 30, 100720

\bibitem[{Hunt(1971)}]{Hunt1}
Hunt, R. 1971, Monthly Notices of the Royal Astronomical Society, 154, 141

\bibitem[{Islam {et~al.}(2020)Islam, Kumar, \& Ghosh}]{Islam1}
Islam, S.~U., Kumar, R., \& Ghosh, S.~G. 2020, Journal of Cosmology and
  Astroparticle Physics, 2020, 030

\bibitem[{Konoplya \& Zinhailo(2020)}]{Roman1}
Konoplya, R.~A., \& Zinhailo, A.~F. 2020, Physics Letters B, 810, 135793

\bibitem[{{Kumar} \& {Ghosh}(2020{\natexlab{a}})}]{Kumar2}
{Kumar}, R., \& {Ghosh}, S.~G. 2020{\natexlab{a}}, \apj, 892, 78

\bibitem[{{Kumar} \& {Ghosh}(2020{\natexlab{b}})}]{Kumar1}
---. 2020{\natexlab{b}}, \jcap, 2020, 053

\bibitem[{{Landau} \& {Lifshitz}(1976)}]{LL1976}
{Landau}, L., \& {Lifshitz}, E. 1976, Mechanics, 1 (Oxford: Pergamon Press)

\bibitem[{{Liu} {et~al.}(2021){Liu}, {Zhu}, \& {Wu}}]{Liu1}
{Liu}, C., {Zhu}, T., \& {Wu}, Q. 2021, Chinese Physics C, 45, 015105

\bibitem[{Liu \& Zhang(2021)}]{Yunlong1}
Liu, Y., \& Zhang, X.~å. 2021, Chinese Physics C

\bibitem[{{Lora-Clavijo} {et~al.}(2015){Lora-Clavijo}, {Cruz-Osorio}, \&
  {Moreno M{\'e}ndez}}]{LoraClavijo2}
{Lora-Clavijo}, F.~D., {Cruz-Osorio}, A., \& {Moreno M{\'e}ndez}, E. 2015,
  \apjs, 219, 30

\bibitem[{{Lora-Clavijo} \& {Guzm{\'a}n}(2013)}]{LoraClavijo1}
{Lora-Clavijo}, F.~D., \& {Guzm{\'a}n}, F.~S. 2013, \mnras, 429, 3144

\bibitem[{L\"u \& Lyu(2020)}]{LuLyu1}
L\"u, H., \& Lyu, H.-D. 2020, Phys. Rev. D, 101, 044059

\bibitem[{MacLeod \& Ramirez-Ruiz(2015)}]{MacLeod1}
MacLeod, M., \& Ramirez-Ruiz, E. 2015, The Astrophysical Journal, 803, 41

\bibitem[{{Narayan} {et~al.}(2012){Narayan}, {S{\"A} dowski}, {Penna}, \&
  {Kulkarni}}]{Narayan1}
{Narayan}, R., {S{\"A} dowski}, A., {Penna}, R.~F., \& {Kulkarni}, A.~K. 2012,
  \mnras, 426, 3241

\bibitem[{{Naveena Kumara} {et~al.}(2020){Naveena Kumara}, {Rizwan}, {Hegde},
  {Sabir Ali}, \& {M}}]{Kumara1}
{Naveena Kumara}, A., {Rizwan}, C.~L.~A., {Hegde}, K., {Sabir Ali}, M., \& {M},
  A.~K. 2020, arXiv e-prints, arXiv:2004.04521

\bibitem[{Ohsugi(2018)}]{Ohsugi1}
Ohsugi, Y. 2018, Astronomy and Computing, 25, 44

\bibitem[{Penner(2011)}]{Penner1}
Penner, A.~J. 2011, Monthly Notices of the Royal Astronomical Society, 414,
  1467

\bibitem[{Penner(2012)}]{Penner2}
---. 2012, Monthly Notices of the Royal Astronomical Society, 428, 2171

\bibitem[{{Roy} \& {Chakrabarti}(2020)}]{Rittick1}
{Roy}, R., \& {Chakrabarti}, S. 2020, \prd, 102, 024059

\bibitem[{{Shaikh} {et~al.}(2021){Shaikh}, {Pal}, {Pal}, \& {Sarkar}}]{Shaikh1}
{Shaikh}, R., {Pal}, K., {Pal}, K., \& {Sarkar}, T. 2021, arXiv e-prints,
  arXiv:2102.04299

\bibitem[{Vincent {et~al.}(2021)Vincent, Wielgus, Abramowicz, Gourgoulhon,
  Lasota, Paumard, \& Perrin}]{Vincent1}
Vincent, F.~H., Wielgus, M., Abramowicz, M.~A., {et~al.} 2021, Astron.
  Astrophys., 646, A37

\bibitem[{{Wei} \& {Liu}(2020)}]{Wei1}
{Wei}, S.-W., \& {Liu}, Y.-X. 2020, arXiv e-prints, arXiv:2003.07769

\bibitem[{{Xu} \& {Stone}(2019)}]{Wenrui1}
{Xu}, W., \& {Stone}, J.~M. 2019, \mnras, 488, 5162

\bibitem[{{Zanotti} {et~al.}(2011){Zanotti}, {Roedig}, {Rezzolla}, \& {Del
  Zanna}}]{Zanotti1}
{Zanotti}, O., {Roedig}, C., {Rezzolla}, L., \& {Del Zanna}, L. 2011, \mnras,
  417, 2899

\bibitem[{{Zhang} {et~al.}(2020){Zhang}, {Wei}, \& {Liu}}]{Zhang1}
{Zhang}, Y.-P., {Wei}, S.-W., \& {Liu}, Y.-X. 2020, Universe, 6, 103

\end{thebibliography}

\end{document}